\documentclass[useAMS,usenatbib,a4paper]{mn2e}
\usepackage{amssymb,amsmath}
\usepackage{graphicx}
\usepackage{natbib}
\usepackage{journal_shortcuts}
\newcommand{\sech}{\mathrm{sech}}
\newcommand{\Sun}{_{\sun}}
\newcommand{\Rot}{_{\mathrm{rot}}}
\newcommand{\Max}{_{\mathrm{max}}}
\newcommand{\Min}{_{\mathrm{min}}}
\newcommand{\Gal}{_{\mathrm{gal}}}
\newcommand{\ICM}{_{\mathrm{ICM}}}
\newcommand{\DM}{_{\mathrm{DM}}}
\newcommand{\Gas}{_{\mathrm{gas}}}
\newcommand{\Stars}{_*}
\newcommand{\Bulge}{_{\mathrm{bulge}}}
\newcommand{\Ram}{_{\mathrm{ram}}}
\newcommand{\degree}{^o}
\newcommand{\Kpc}{\,\textrm{kpc}}
\newcommand{\Mpc}{\,\textrm{Mpc}}
\newcommand{\PC}{\,\textrm{pc}}
\newcommand{\CM}{\,\textrm{cm}}
\newcommand{\Myr}{\,\textrm{Myr}}
\newcommand{\Kms}{\,\textrm{km}\,\textrm{s}^{-1}}
\newcommand{\Erg}{\,\textrm{erg}}
\newcommand{\ccm}{\,\textrm{cm}^{-3}}
\newcommand{\gccm}{\,\textrm{g}\,\textrm{cm}^{-3}}

\title[Wakes of ram pressure stripped disc galaxies]%
{Wakes of ram pressure stripped disc galaxies}

\author[E. Roediger, M. Br\"uggen and M.~Hoeft]%
{Elke Roediger%
\thanks{E-mail:
e.roediger@iu-bremen.de; m.brueggen@iu-bremen.de; m.hoeft@iu-bremen.de}
and  
Marcus Br\"uggen\footnotemark[1]
and  
Matthias Hoeft\footnotemark[1]%
\\
International University Bremen, P.O. Box 750\,561, 28725 Bremen,
Germany}

\begin{document}

\date{Accepted. Received; in original form }

\pagerange{\pageref{firstpage}--\pageref{lastpage}} \pubyear{2006}

\maketitle

\label{firstpage}

\begin{abstract}
Spiral galaxies that move through the intracluster medium lose a substantial
amount of their gas discs due to ram pressure stripping. The recent
observations of NGC 4388 by \citet{oosterloo05} reveal a tail of stripped gas
of $\sim 100\Kpc$ behind the source galaxy.  We present first 3D
hydrodynamical simulations of the evolution of such ram pressure stripped
tails.

We find that if the ICM wind does not vary significantly over a period of a
few $100\Myr$, subsonic galaxies produce a tail with regular features similar
to a von-Karman vortex street. In this case, the tail widens systematically by
about $45\Kpc$ per $100\Kpc$ distance behind the source galaxy. The widening
rate is independent of the galaxy's inclination for a large range of
inclinations. For supersonic galaxies, the tail is more irregular than for
subsonic ones.

The tail observed for NGC~4388 is narrower than the tails in our simulations.
Reasons for this difference may be additional physical processes such as heat
conduction or viscosity. In addition, we conclude that the observed S-shape of
this tail is not due to von Karman oscillations, because this galaxy is likely
to move supersonically. A reason for the observed shape may be motions in the
ambient ICM. 

Finally, we discuss implications for the distribution of metals in the ICM due
to ram pressure stipping. 
\end{abstract}

\begin{keywords}
galaxies: spiral -- galaxies: evolution -- galaxies: ISM -- intergalactic medium
\end{keywords}

%
%
%
%
%
\section{Introduction}
%
Ram pressure stripping, i.e. the removal of a galaxy's gas disc due to its
motion through the intracluster medium (ICM), has been studied mainly
with regard to the effect on the galaxy
(e.g. \citealt{abadi99,quilis00,schulz01,vollmer01a,marcolini03,roediger05,roediger06}).  It has been shown that the interaction between the
ICM and the galactic interstellar medium (ISM) can remove a significant amount
of gas from galaxies near cluster centres. Thus, ram pressure
stripping is a good candidate to explain observations such as the HI deficiency
of cluster disc galaxies (see e.g.~\citealt{cayatte90,cayatte94,solanes01})
and the truncated star forming discs in the Virgo cluster
(\citealt{koopmann98,koopmann04b,koopmann04a,koopmann05}). A distinct feature
of ram pressure stripped galaxies is that their gas discs are distorted or
truncated while their stellar discs appear undisturbed, because ram pressure
stripping affects only the gaseous components of the galaxy. Such examples
have been observed, e.g.~NGC 4522 (\citealt{kenney99,kenney01,kenney04},
\citealt{vollmer04a}), NGC 4548 (\citealt{vollmer99}), NGC 4848
(\citealt{vollmer01}). In addition to the effect on the galaxy, ram pressure
stripping plays an important role in the evolution of the ICM because
the gas lost by galaxies enriches the ICM with metals (see
e.g. \citealt{schindler05,domainko05} and references therein).

A further example of ram pressure stripped galaxies is the Virgo spiral NGC
4388. It has a small distance to the cluster centre and a high radial velocity
($\sim 1400\Kms$) with respect to the cluster mean (see \citealt{vollmer03a}
and references therein). There is strong evidence that it has suffered
ram pressure stripping in its recent history. It was observed to be HI
deficient by \citet{cayatte90}. Moreover, both, ionised and neutral gas have
been found at a few kpc distance (also see \citealt{vollmer03a} for further
references). \citet{vollmer03a} reported the detection of atomic gas at a
projected distance of more than $20\Kpc$ from the galactic centre. This
extraplanar gas is interpreted as the tail of gas stripped from NGC
4388. On the basis of sticky particle simulations, \citet{vollmer03a}
concluded that the main morphological features and velocity structure of NGC
4388 and the extragalactic gas clouds can be explained by ram pressure
stripping.

Recently, \citet{oosterloo05} presented observations of a $\sim120\Kpc$ long
tail of HI gas associated with NGC 4388, and also suggest that this tail is
due to ram pressure stripping of this galaxy, either in the ICM of the Virgo
cluster or in the halo of the nearby elliptical galaxy M86. Two prominent
features of the tail are its flaring width and its slight S-shape.

Although ram pressure stripping has been studied by several
groups, only very few investigations of the resulting gas tails are
available. \citet{stevens99} and \citet{acreman03} simulated ram pressure
stripping of hot gas halos from spherical galaxies. They predicted that bow
shocks and a tail of hot material will be observable in X-rays. Both groups
used cylindrical 2D hydro codes for their simulations.

To our knowledge, so far no simulations of tails of cool gas stripped from
disc galaxies have been presented. However, the physics that determines the
dynamics of such tails is complex and interesting:
\begin{itemize}
\item According to basic hydrodynamics, any object that moves through a fluid
  produces a wake. The characteristics of this wake depend on the Reynolds
  number (see any hydrodynamics text book, e.g.~\citealt{batchelor}). For
  small Reynolds numbers ($\mathrm{Re} \lesssim 50$), the wake is thin. For
  moderate Reynolds numbers ($\mathrm{Re} \sim 100$), features similar to the
  von Karman vortex street develop. The wake's width can reach several times
  the width of the body. For high Reynolds numbers ($\mathrm{Re} \sim 1000$),
  the wake becomes turbulent. Especially for moderate and high Reynolds
  numbers, the wake widens significantly with increasing distance to the
  object. \newline
  The value of the Reynolds number in the ICM is a matter of ongoing debate
  (see e.g.~\citealt{reynolds05}), with a large range of values suggested.
\item The slight S-shape of the tail of NGC 4388 present in the observation of
  \citet{oosterloo05} is reminiscent of a von Karman vortex street. However,
  classical von Karman vortex streets are usually generated behind ``2D
  bodies'', like cylinders, that force the flow into a translational symmetry
  along the axis of the body. In contrast, the flow past a galaxy is a full 3D
  flow and thus possesses a more complex wake structure.
\item The stripping process alone may lead to a widening of the galaxy's tail
  because the ICM is forced to flow around the galactic gas disc. Therefore,
  the galactic gas can aquire velocity components perpendicular to the main ICM
  wind direction. 
\item Additions to the dynamics in the galactic wake/tail could arise from the
  rotation of the disc gas. As the stripped galactic gas moves away from the
  galaxy, the radial gravitational acceleration ceases to operate on the
  stripped gas. In the extreme case of a face-on galaxy where the radial
  acceleration is simply switched off, a stripped gas package would move
  linearly with $v=v\Rot$ perpendicular to $\vec{v}\ICM$ from that moment
  on. The ICM wind can accelerate the stripped gas package to maximal $v\ICM$
  in wind direction. The superposition of a motion with $v\ICM$ in wind
  direction and $v\Rot$ in any direction perpendicular to this would lead to a
  widening angle of $\alpha=2\, \arctan(v\Rot/v\ICM)$. This process should be
  inclination-dependent. Of course, stripped gas has to pass a transition
  region where the radial gravitational acceleration ceases to
  operate.
\item The inclination angle between the galactic rotation axis and the
  galaxy's direction of motion may influence the wake due to geometrical
  reasons alone, i.e. the effective cross-section of the ``obstacle'' in the
  ICM wind changes. Furthermore, the flow patterns of subsonic and supersonic
  cases are expected to differ because in supersonic cases a bow shock
  develops.
\end{itemize}
Most of the aspects listed above require a 3D treatment of the problem.
Here, we present first 3D hydrodynamical adaptive-mesh simulations of the
evolution of tails of ram pressure stripped galaxies. 
%

\section{Method}
%
We study the motion of the galaxy through the cluster in the
rest frame of the galaxy. Therefore, its motion translates into an ICM wind
flowing past the galaxy.

\subsection{Code} \label{sec:code}
The simulations were performed with the FLASH code (\citealt{fryxell00}), a
multidimensional adaptive mesh refinement code. 
It solves the Riemann problem
on a Cartesian grid using the Piecewise-Parabolic Method (PPM), which is
particularly well suited to resolve shocks. The gas obeys a polytropic
equation of state with an adiabatic index of $\gamma=5/3$. 
All boundaries but the inflow boundary are open. The simulations presented
here are performed in 3D. In order to follow the evolution of the wake, we use
a large simulation box of size $(x\Min,x\Max)\times (y\Min,y\Max)\times
(z\Min,z\Max)=(-162\Kpc,162\Kpc)\times (-64\Kpc,260\Kpc)\times
(-162\Kpc,162\Kpc)$ for subsonic cases and $(-121.5\Kpc,121.5\Kpc)\times
(-64\Kpc,260\Kpc)\times (-121.5\Kpc,121.5\Kpc)$ for supersonic ones. The
galactic centre is located at $(x\Gal,y\Gal,z\Gal)=(0,0,0)$. The ICM wind is
flowing along the $y$-axis into the positive direction. We use 6 levels of
refinement, which leads to an effective resolution of $316.5\PC$ and an
effective number of grid cells of $(1024)^3$ for the subsonic cases.  However,
at a distance of more than $64\Kpc$ behind the galaxy (behind=downstream) we
limit the effective resolution to $633\PC$ (5 levels of refinement) to limit
computational costs. For the same reason, the number of refinement levels is
restricted to 5 for the supersonic runs, where a multitude of shocks and waves
causes massive refinement (see Fig.~\ref{fig:denscut_supersonic}). The
refinement criteria are the standard density and pressure gradient
criteria. The galactic disc is always refined.

In order to be able to identify the galactic gas after it is stripped from the
galaxy, we use a ``dyeing'' technique: The FLASH code offers the opportunity
to advect mass scalars along with the density. We utilise one of these mass
scalars, $f$, to contain the fraction of galactic gas in each cell. Initially
this array has the value 1 in the region of the galactic disc and 0
elsewhere. As a result, every cell where $f>0$ contains a certain amount
of gas that has originally been inside the galaxy. At each timestep, the
quantity $f\rho$ gives the local density of galactic gas.

\subsection{Model galaxy}
\begin{figure}
\centering\resizebox{0.7\hsize}{!}{\includegraphics[angle=0]{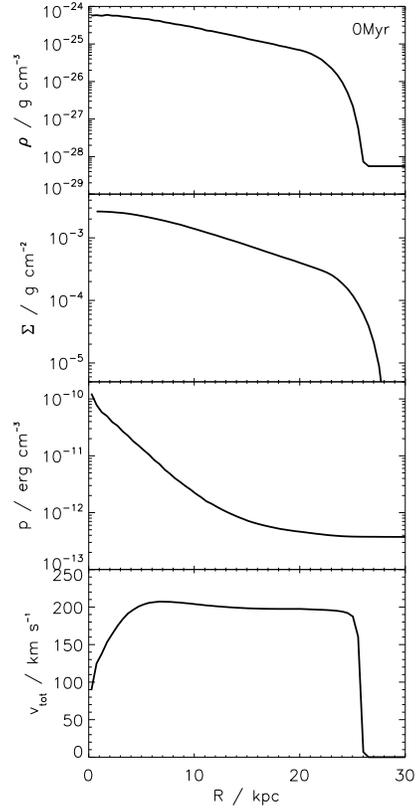}}
\caption{Radial profiles of the density, $\rho$, pressure, $p$,
  and rotation velocity, $v\Rot$, in the galactic plane for the initial
  model. Also the radial profile for the projected ISM surface density,
  $\Sigma$, is shown.}
\label{fig:initial_profiles}
\end{figure}

We model a massive spiral galaxy with a flat rotation curve at $200\Kms$. It
consists of a gas disc, a stellar disc, a stellar bulge and a dark matter (DM)
halo. For the gravitational potential of the stellar disc, bulge and DM halo,
we use the following analytical descriptions:
\begin{description}
\item[Stellar disc:] Plummer-Kuzmin disc, see \citet{miyamoto75} or
  \citet{binneytremaine}. Such discs are characterised by their mass,
  $M\Stars$, and radial and vertical scale lengths, $a\Stars$ and $b\Stars$,
  respectively.
\item[Stellar bulge:] Spherical Hernquist bulge (see \citealt{hernquist93}). In
  case of a spherical bulge, the gravitational potential, $\Phi$, depends on
  radius, $r$, as $\Phi(r)= -
  \frac{G\,M\Bulge}{r+r\Bulge}$, where $M\Bulge$ is the mass of the bulge,
  $r\Bulge$ the scale radius and $r$ the spherical radius.
\item[DM halo:] The spherical model of \citet{burkert95}, including the
  self-scaling relations, i.e. the DM halo is characterised by the radial
  scale length, $r\DM$, alone. For the equation of the analytical potential see
  also \cite{mori00}.
\end{description}
The self-gravity of the gas is neglected as the gas contributes only a small
fraction to the overall galactic mass. The parameters of our galaxy model are
summarised in Table~\ref{tab:galaxy_parameters}.
%
\begin{table}
\caption{Galaxy model parameters.}
\label{tab:galaxy_parameters}
\centering\begin{tabular}{lll}
\hline
             &$M\Stars$   & $10^{11}M\Sun$ \\
stellar disc &$a\Stars$   & 4\,kpc                    \\
             &$b\Stars$   & 0.25\,kpc                 \\
\hline
bulge        &$M\Bulge$   & $10^{10}M\Sun$ \\
             &$r\Bulge$   & 0.4\,kpc                  \\
\hline
DM halo      &$r\DM$      & 23\,kpc                   \\
\hline
             &$M\Gas$     & $10^{10}M\Sun\,^*$ \\
gas disc     &$a_{\Sigma}$& 7\,kpc                     \\
             &$b\Gas$  & 0.4\,kpc       \\
             &$v\Rot$     & 200$\Kms$             \\
\hline
\end{tabular}
\end{table}
In order to prevent steep density gradients in the galactic plane and in the
galactic centre, the gas disc is described by a softened exponential disc :
\begin{equation}
\rho(R,Z)=\frac{M\Gas}{2\pi a\Gas^2 b\Gas}  \,0.5^2\,\sech\left(\frac{R}{a\Gas}\right)\, \sech\left(
\frac{|Z|}{b\Gas}\right) \label{eq:dens_exp_soft}.
\end{equation}
The coordinates $(R,Z)$ are galactic cylindrical coordinates. The radial and
vertical scale lengths are $a\Gas$ and $b\Gas$, respectively. For $R\gtrsim
a\Gas$ and $|Z|\gtrsim b\Gas$, this density distribution converges towards the
usual exponential disc $\rho(R,Z)=\frac{M\Gas}{2\pi a\Gas^2 b\Gas}
\exp(-R/a\Gas)\,\exp(-|Z|/b\Gas)$. For the corresponding exponential disc,
$M\Gas$ is the total gas mass. We chose $M\Gas$ such that in the outer regions
the gas disc converges to an exponential gas disc with 10\% of the stellar
disc mass, i.e.  $M\Gas=0.1 M\Stars$. The integrated gas mass amounts to
$6\cdot 10^{9}M\Sun$. Given the density distribution in the disc, its pressure
and temperature distribution are set such that hydrostatic equilibrium with
the ICM is maintained in the direction perpendicular to the disc plane. In
radial direction, the disc's rotation velocity is set so that the centrifugal
force balances the gravitational force and pressure gradients. We have cut the
gas disc smoothly to a finite radius of $26\Kpc$ by multiplying the 
density distribution $\rho(R,Z)$ with $0.5[1+ \cos(\pi(R-20\Kpc)/6\Kpc ) ]$
for $20\Kpc < R \le 26\Kpc$. Figure~\ref{fig:initial_profiles} shows radial
profiles of density, surface density, pressure and rotation velocity for the
initial model.
%
%

We measure the inclination angle, $i$, of the galaxy as the angle between
its rotation axis and the ICM wind direction. Thus $0\degree$ corresponds to
to a face-on motion of the galaxy, and $90\degree$ to edge-on.

\subsection{ICM conditions}
For a first experiment, we expose model galaxies to constant ICM
winds.  We list $p\Ram$, $v\ICM$, 
$\rho\ICM$ and inclination angle, $i$, in Table~\ref{tab:wind_parameters}. 
%
\begin{table}
\caption{ICM wind parameters. Ram pressure $p\Ram$, ICM wind density
  $\rho\ICM$, ICM wind velocity $v\ICM$, inclination angle $i$.}
\label{tab:wind_parameters}
\centering\begin{tabular}{llll}
\hline
$p\Ram/$            & $\rho\ICM/$ & $v\ICM/$ &             \\
$(\Erg\ccm)$        & $(\gccm)$   & $(\Kms)$ & $i/\degree$ \\
\hline
$6.4\cdot 10^{-11}$ & $10^{-26}$  & 800      & 30 \\
$6.4\cdot 10^{-11}$ & $10^{-27}$  & 2530     & 30   \\
$6.4\cdot 10^{-12}$ & $10^{-27}$  & 800      & 30, 75 \\
$6.4\cdot 10^{-12}$ & $10^{-28}$  & 2530     & 30 \\
\hline
\end{tabular}
\end{table}
%
We chose the ICM temperature such that the sound speed is $1000\Kms$. We refer
to the two ram pressures as medium ($p\Ram=6.4\cdot 10^{-12}\Erg\ccm$) and
strong ($p\Ram=6.4\cdot 10^{-11}\Erg\ccm$) ram pressure. For both ram
pressures, we test a subsonic ($v\ICM=800\Kms$, Mach number 0.8) and a
supersonic ($v\ICM=2530\Kms$, Mach number 2.53) case.

For the subsonic runs, we start with the ICM at rest and then increase the
inflow velocity during the first $50\Myr$. For the supersonic runs, this
procedure takes too long until a true supersonic flow including a bow shock
develops. Therefore, for supersonic runs, we set $v\ICM$ everywhere in the ICM
already initially. In this procedure, the simulations starts a bit violently,
but after a bit less than $100\Myr$ a true bow shock has developed. This
difference in initialisation leads to a temporal offset between subsonic and
supersonic runs. A time $t$ in a supersonic run corresponds to approximately
$t+50\Myr$ in a subsonic run.

\section{Results}
%
\subsection{Snapshots in slices}
\begin{figure}
\centering\resizebox{\hsize}{!}%
{\includegraphics[trim=20 60 0 -20,clip,angle=90]{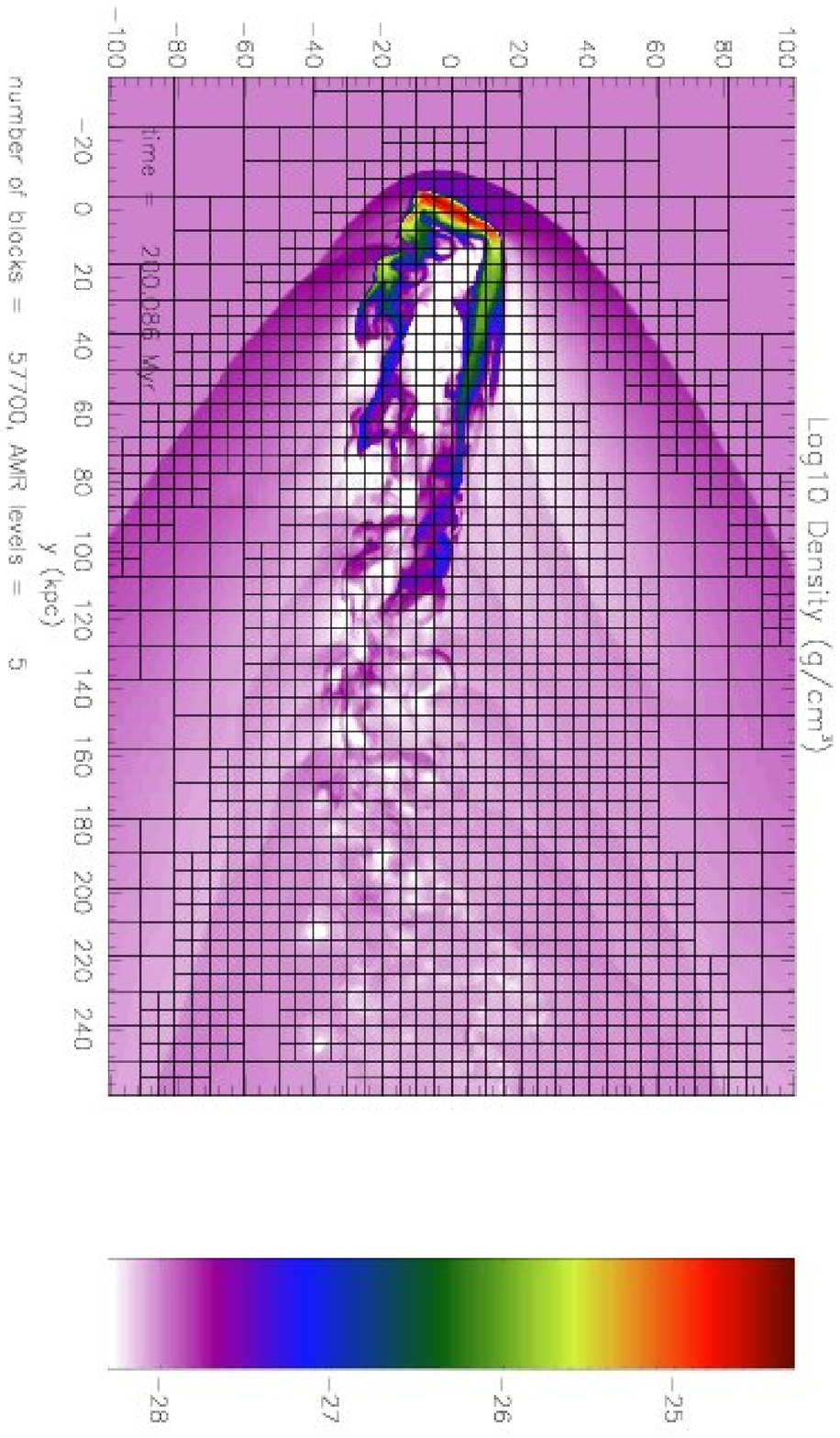}
}
\centering\resizebox{\hsize}{!}%
{
\includegraphics[trim=20 60 0 -20,clip,angle=90]{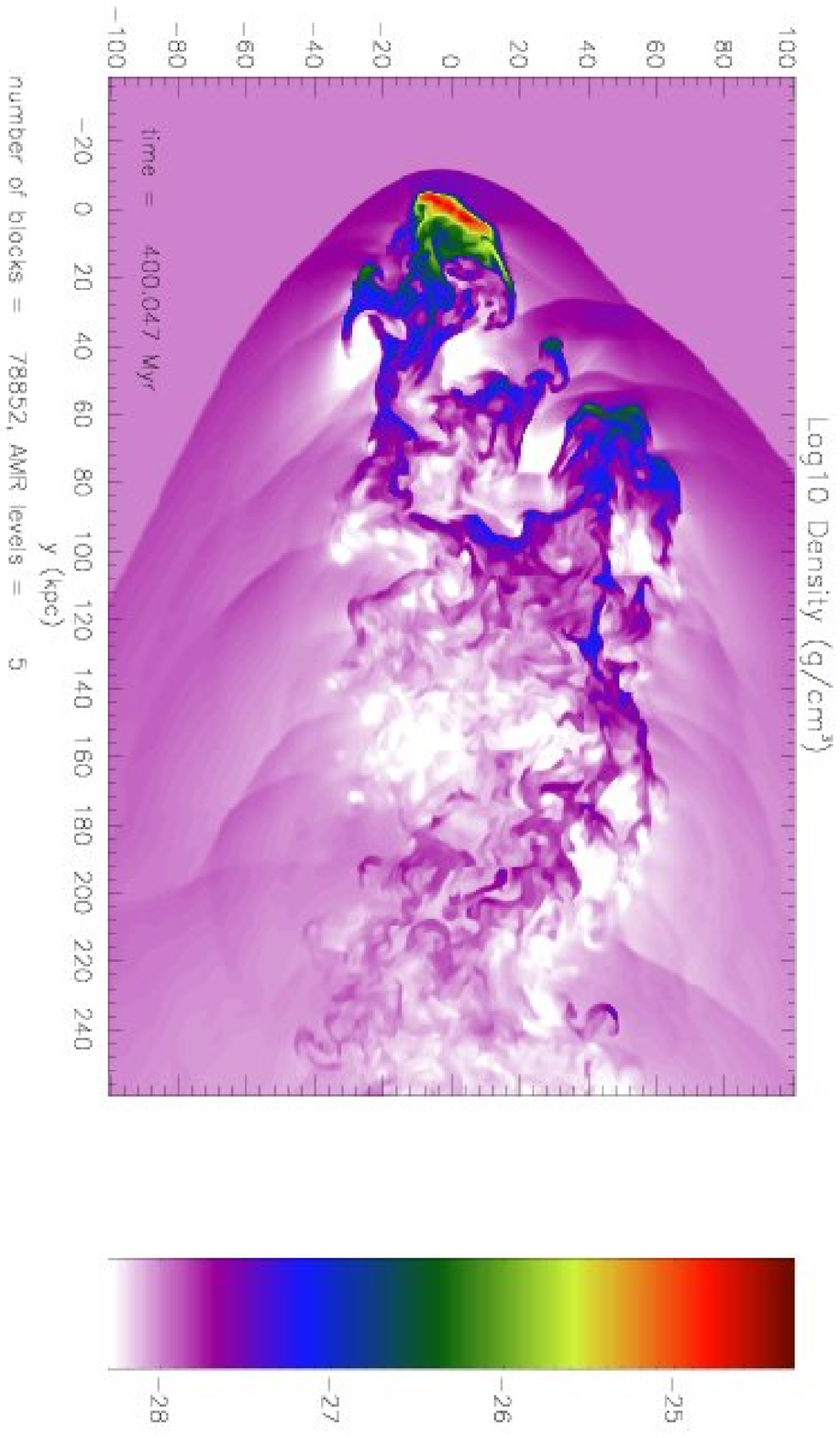}
}
\centering\resizebox{\hsize}{!}%
{\includegraphics[trim=20 60 0 -20,clip,angle=90]{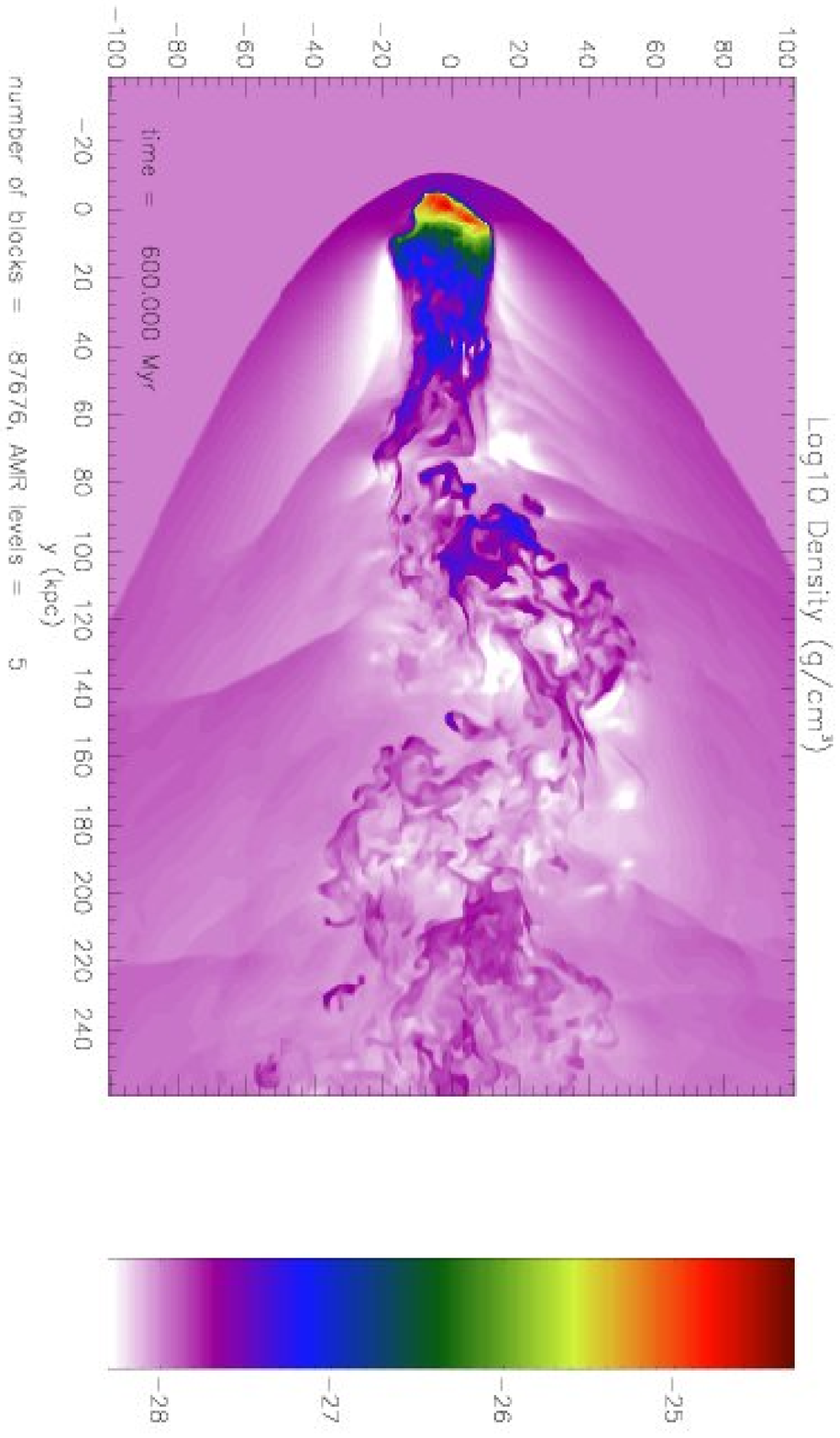}
}
\caption{Slice through simulation box in the $y$-$z$-plane for the supersonic
  run with medium ram-pressure and $i=30\degree$. The local gas density is
  shown colour-coded for different time steps. The refinement is demonstrated
  in the first plot, one square shows the size of one block$=8^3$ grid cells.}
\label{fig:denscut_supersonic}
\end{figure}

Figure~\ref{fig:denscut_supersonic} shows the gas density distribution in a
slice through the simulation box parallel to the $y$-$z$-plane at $x=0$ for
different timesteps for the case of medium ram-pressure and supersonic ICM
flow.
%
%
In addition to the bow shock, the stripped material induces a rich shock
structure in the region behind the bow shock. The stripped material
fragments strongly due to turbulence and Rayleigh-Taylor-instability.

\subsection{Projected gas densities}
\begin{figure}
\centering\resizebox{\hsize}{!}%
{\includegraphics{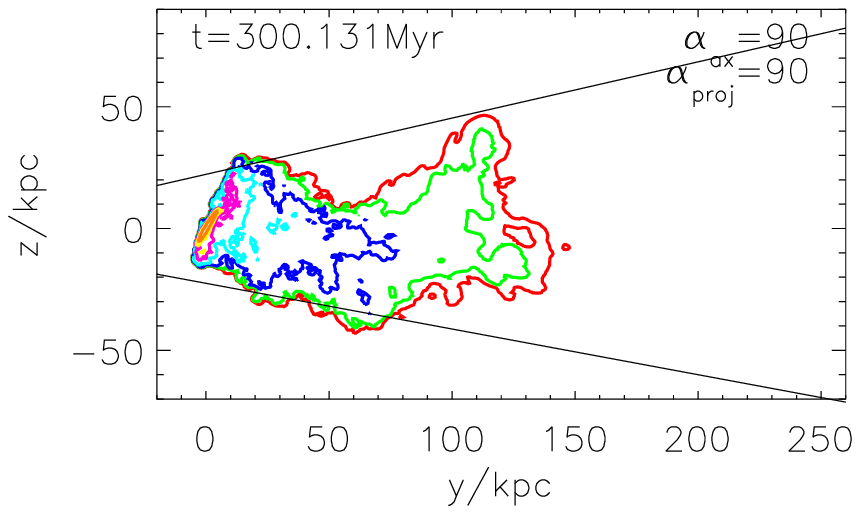}
\includegraphics{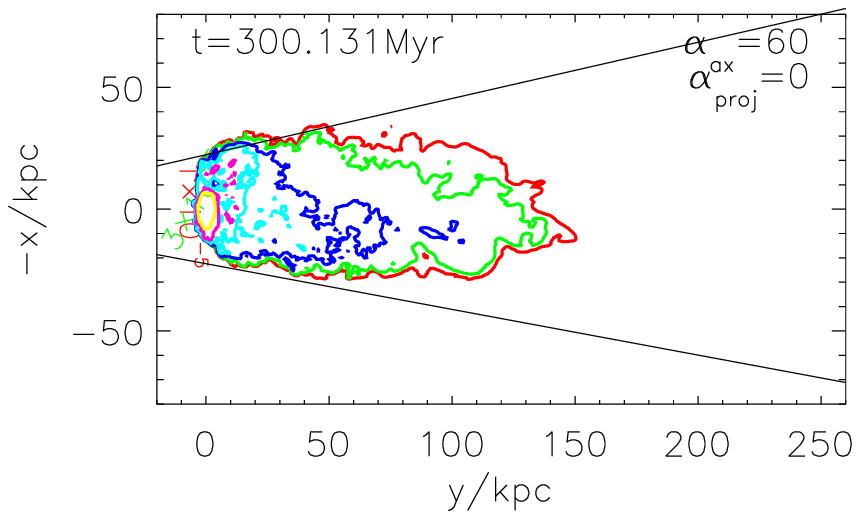}}
\centering\resizebox{\hsize}{!}%
{\includegraphics{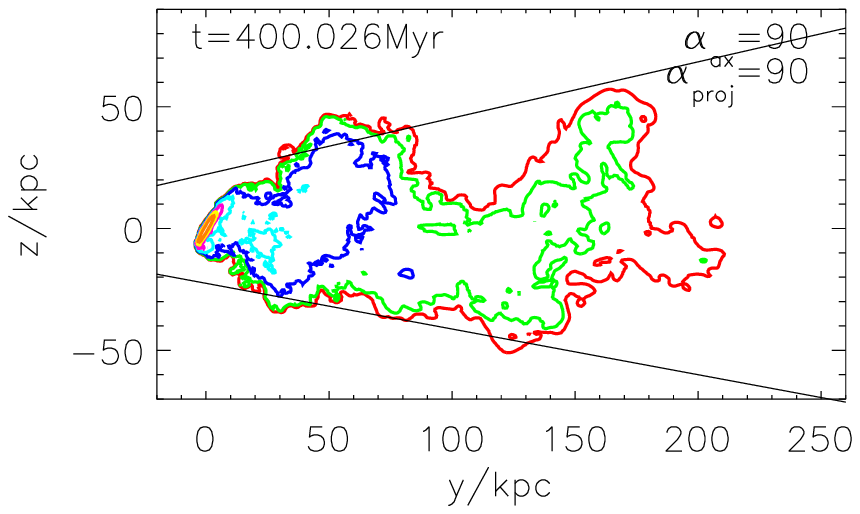}
\includegraphics{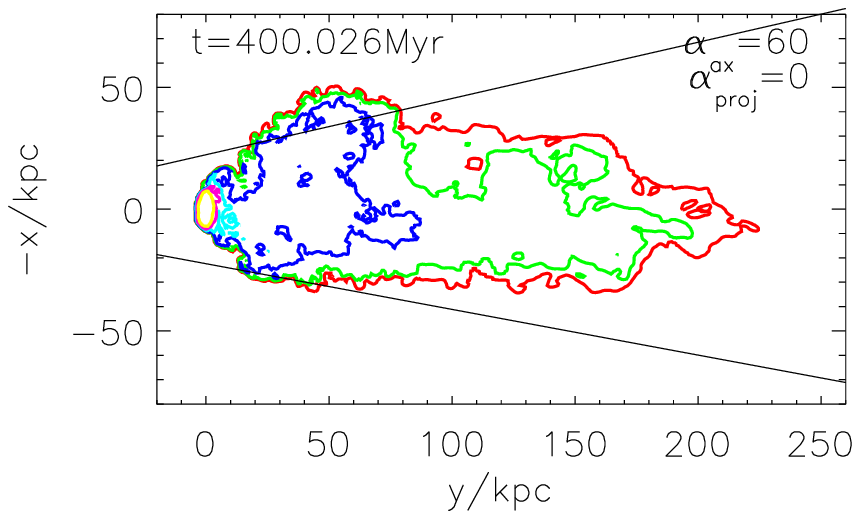}}
\centering\resizebox{\hsize}{!}%
{\includegraphics{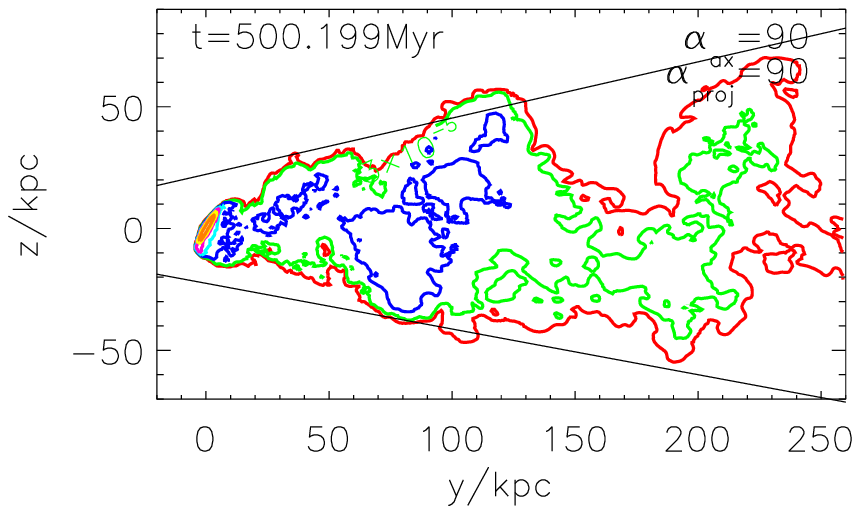}
\includegraphics{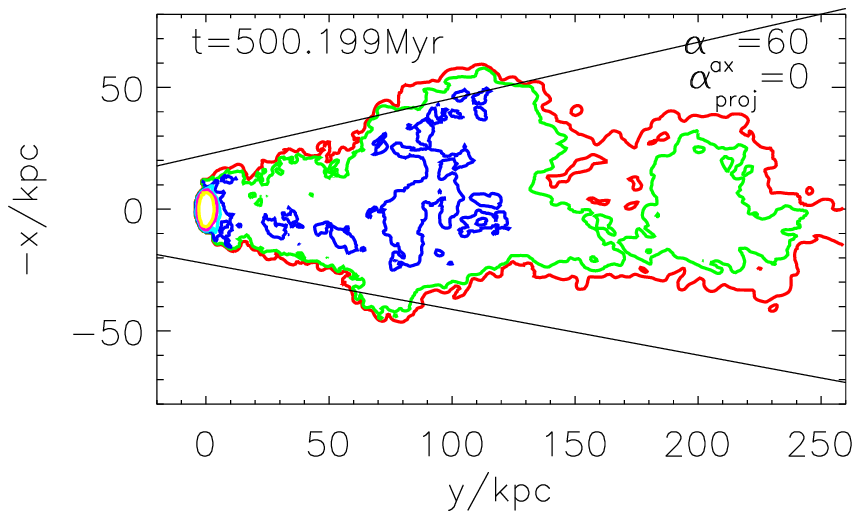}}
\centering\resizebox{\hsize}{!}%
{\includegraphics{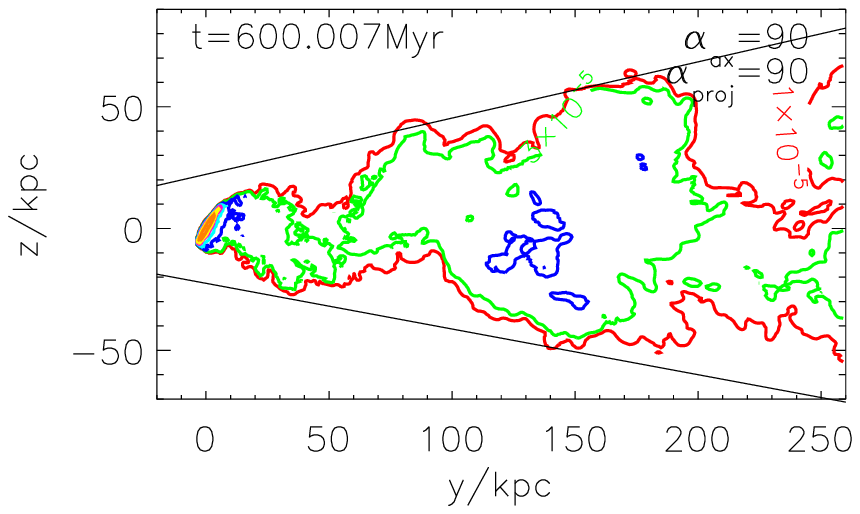}
\includegraphics{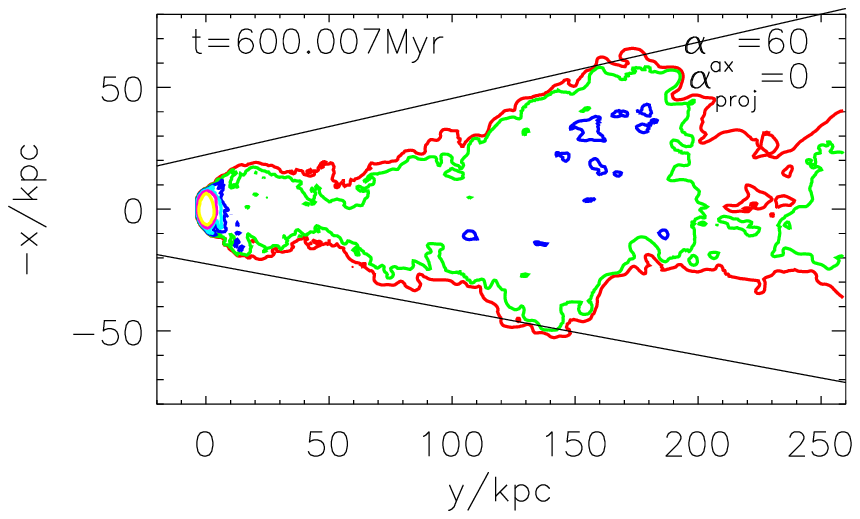}}
\centering\resizebox{\hsize}{!}%
{\includegraphics{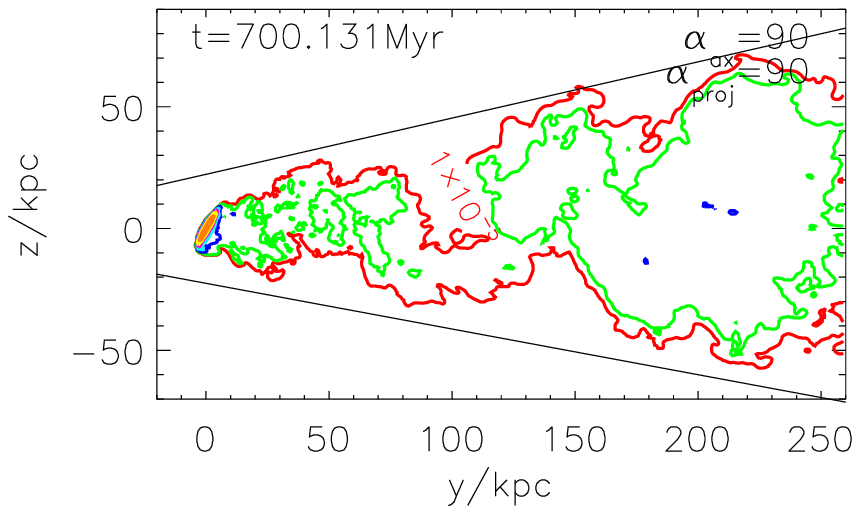}
\includegraphics{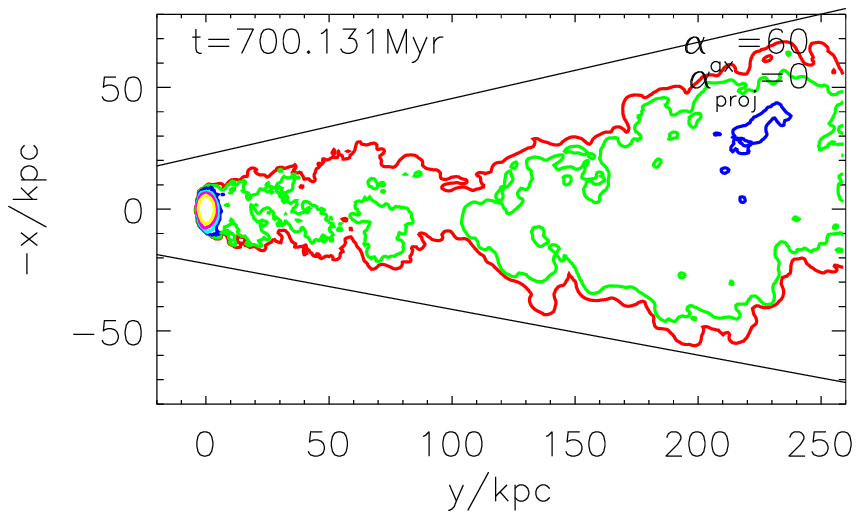}}
\caption{Projected galactic gas density for medium ram-pressure, Mach number
  0.8, $i=30\degree$. Lowest contour is $10^{-5}\mathrm{g}\,\CM^{-2} \hat{=}
  6.0\cdot 10^{18}\,\CM^{-2}$. The contour spacing is half an order of
  magnitude. The left column is for projection along the $x$-axis, the right
  column for projection along the $z$-axis. The time of each panel is denoted
  in its upper left corner. The two numbers in the upper right corner denote
  the projection direction with respect to the galaxy: $\alpha_{\mathrm{ax}}$
  is the angle between the line-of-sight and the galactic rotation axis. If
  the line-of-sight is projected into the grid's $x$-$z$-plane, the resulting
  line has an angle of $\alpha_{\mathrm{proj}}$ with the negative part of the
  grid's $z$-axis. $\alpha_{\mathrm{ax}}$ and $\alpha_{\mathrm{proj}}$ are
  given in degree.}
\label{fig:sdens_subsonic}
\end{figure}

\begin{figure}
\centering\resizebox{\hsize}{!}%
{\includegraphics{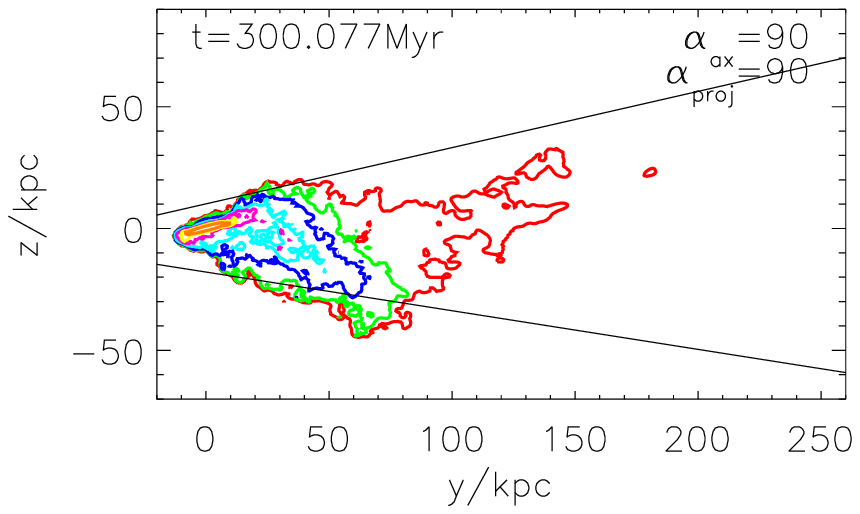}
\includegraphics{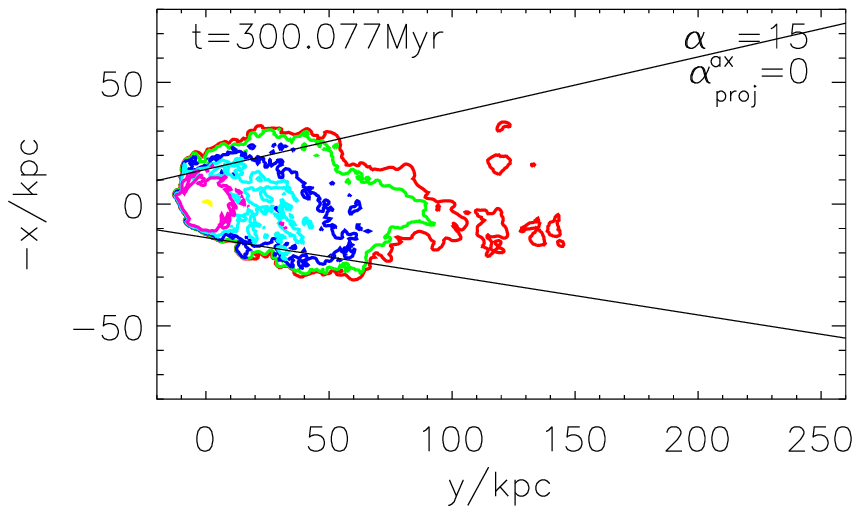}}
\centering\resizebox{\hsize}{!}%
{\includegraphics{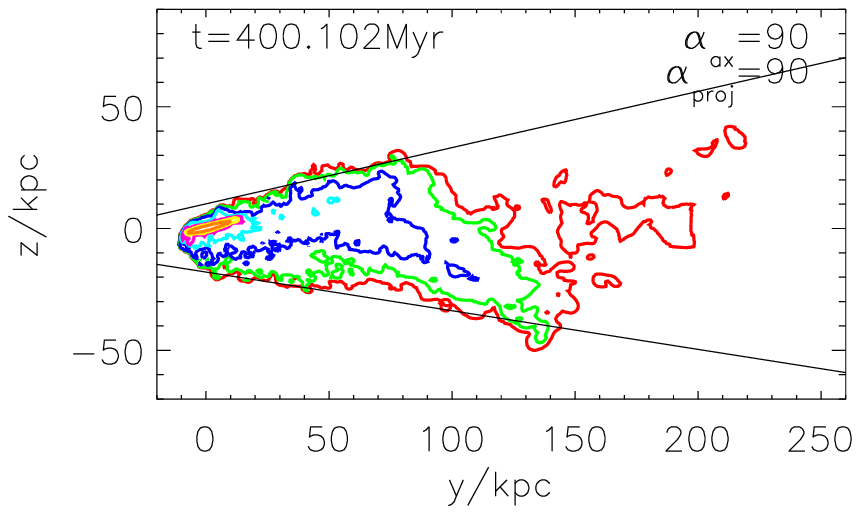}
\includegraphics{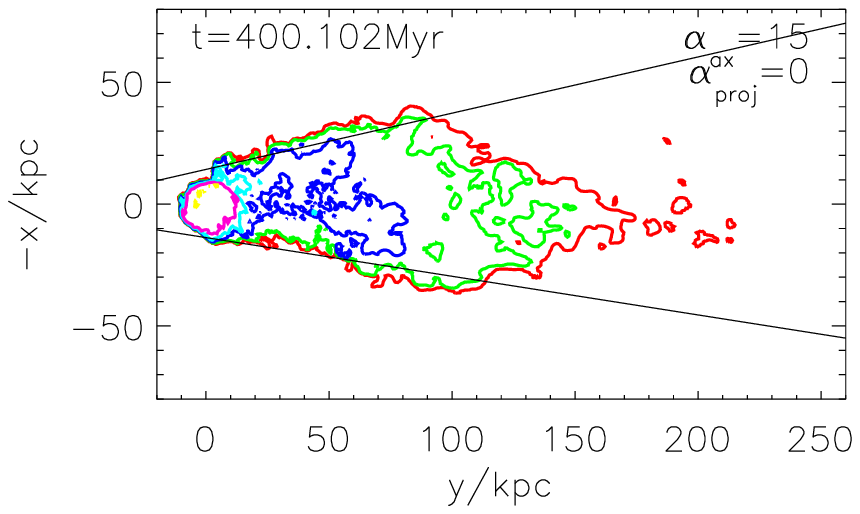}}
\centering\resizebox{\hsize}{!}%
{\includegraphics{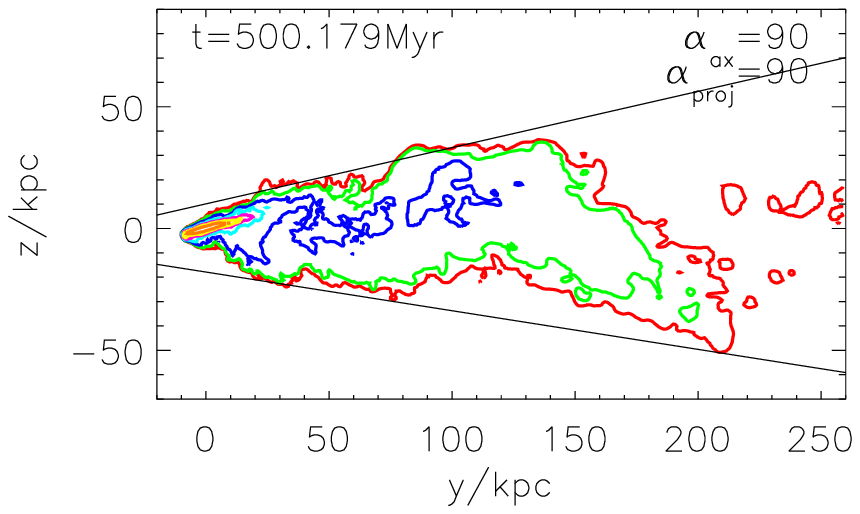}
\includegraphics{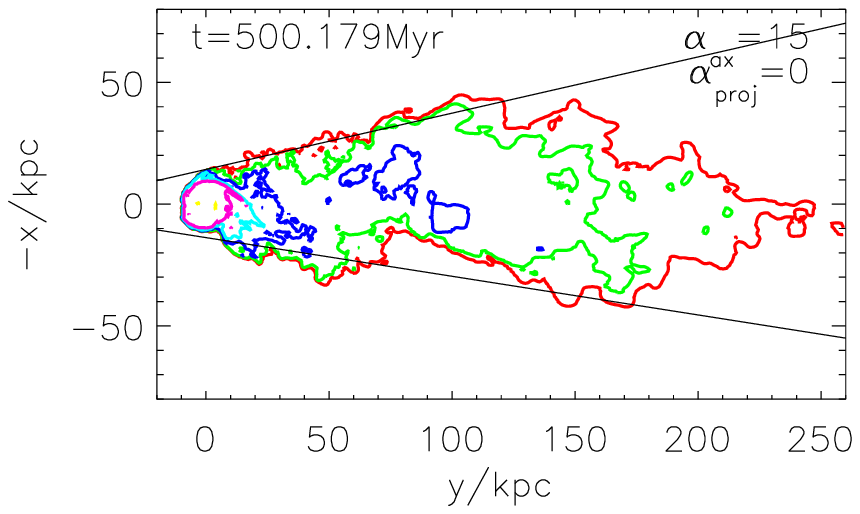}}
\centering\resizebox{\hsize}{!}%
{\includegraphics{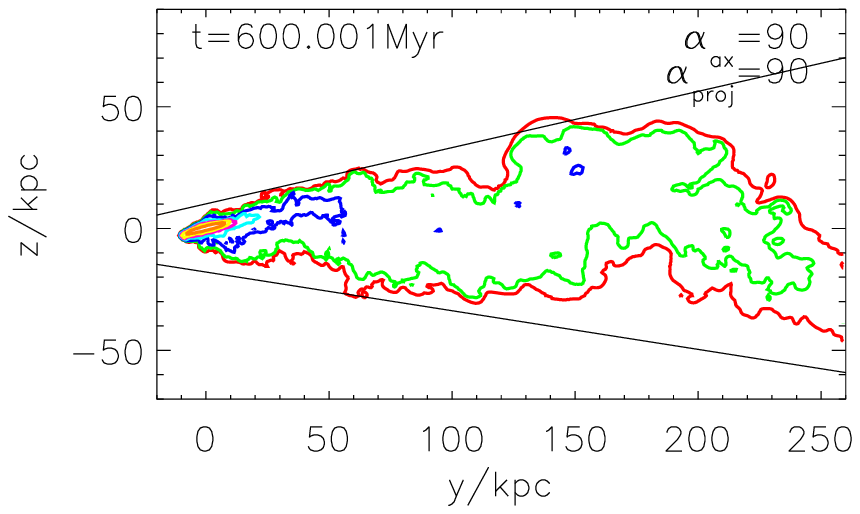}
\includegraphics{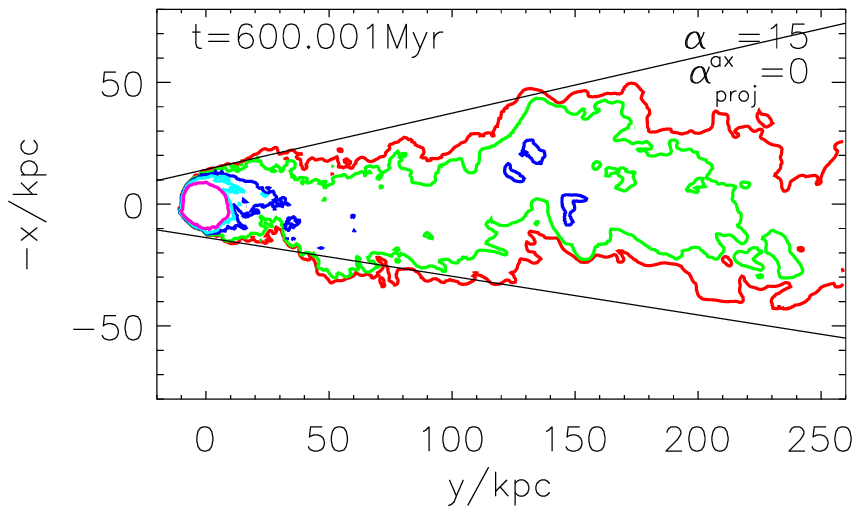}}
\centering\resizebox{\hsize}{!}%
{\includegraphics{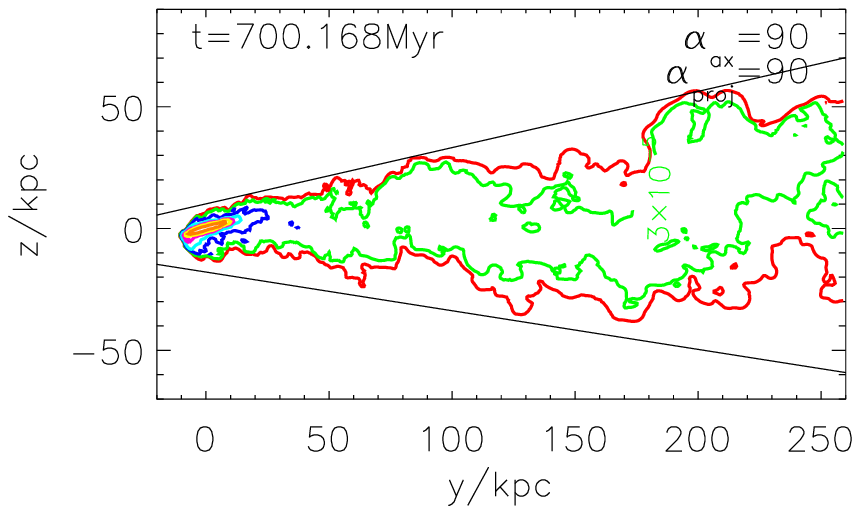}
\includegraphics{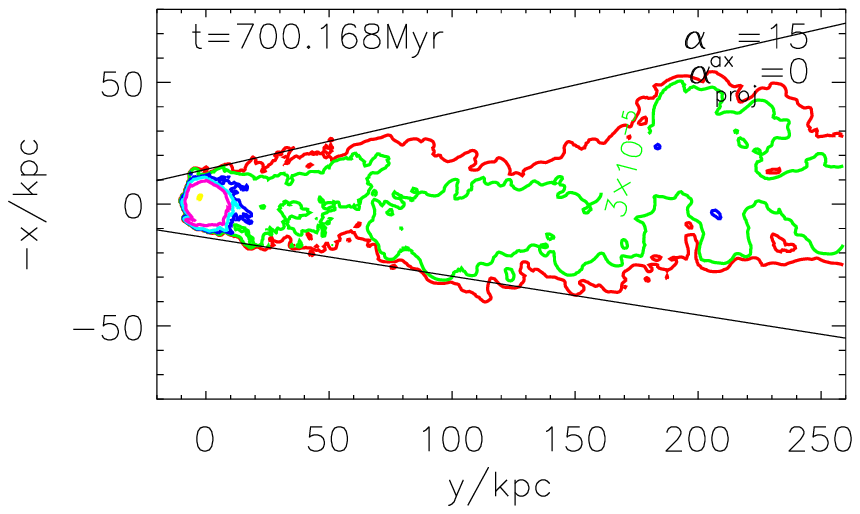}}
\caption{Projected galactic gas density for medium ram-pressure, Mach number
  0.8, $i=75\degree$. Otherwise same as Fig.~\ref{fig:sdens_subsonic}.}
\label{fig:sdens_subsonicI75}
\end{figure}

\begin{figure}
\centering\resizebox{\hsize}{!}%
{\includegraphics{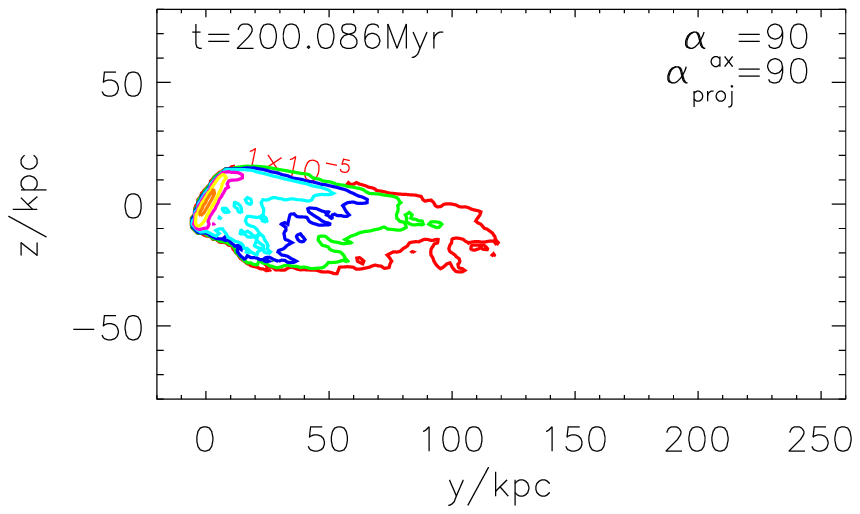}
\includegraphics{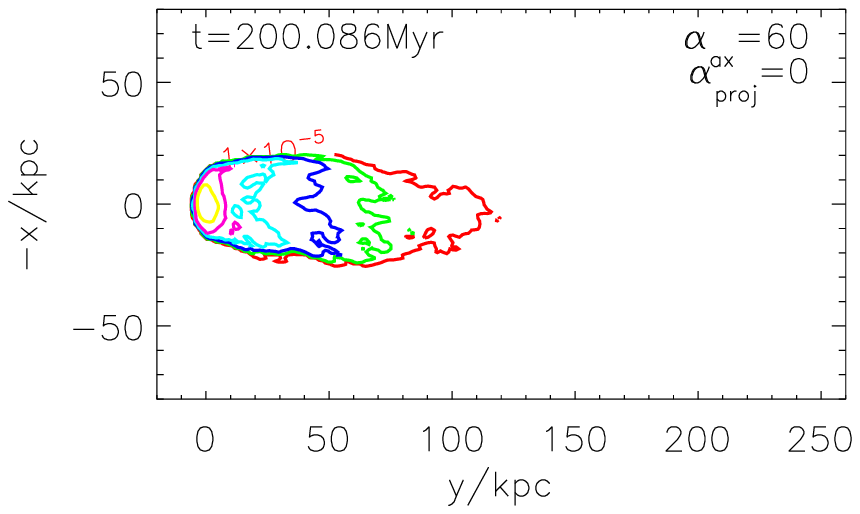}
}
\centering\resizebox{\hsize}{!}%
{\includegraphics{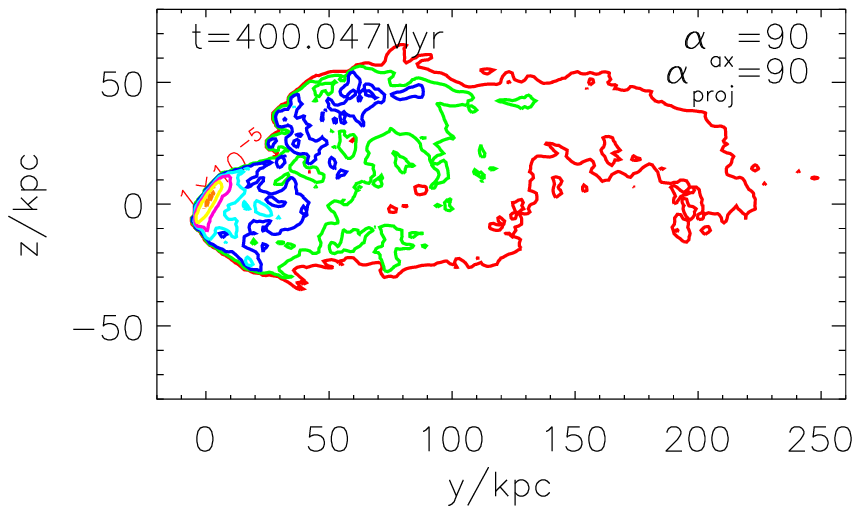}
\includegraphics{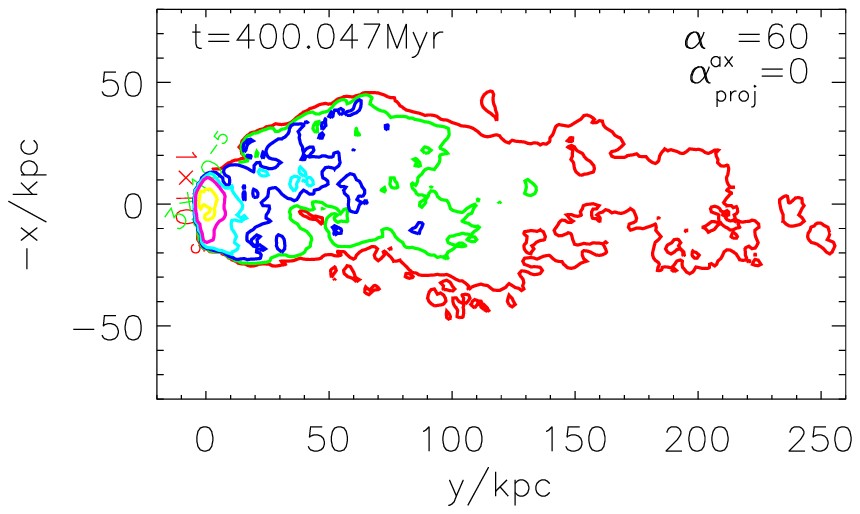}
}
\centering\resizebox{\hsize}{!}%
{\includegraphics{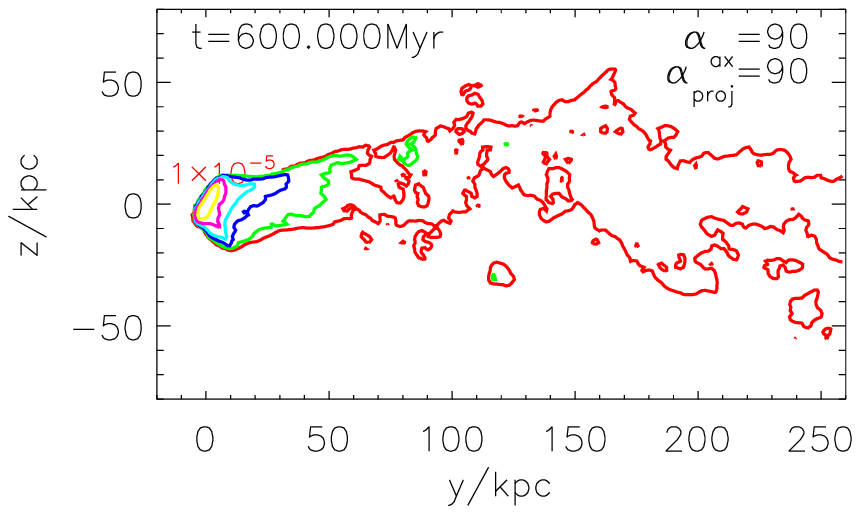}
\includegraphics{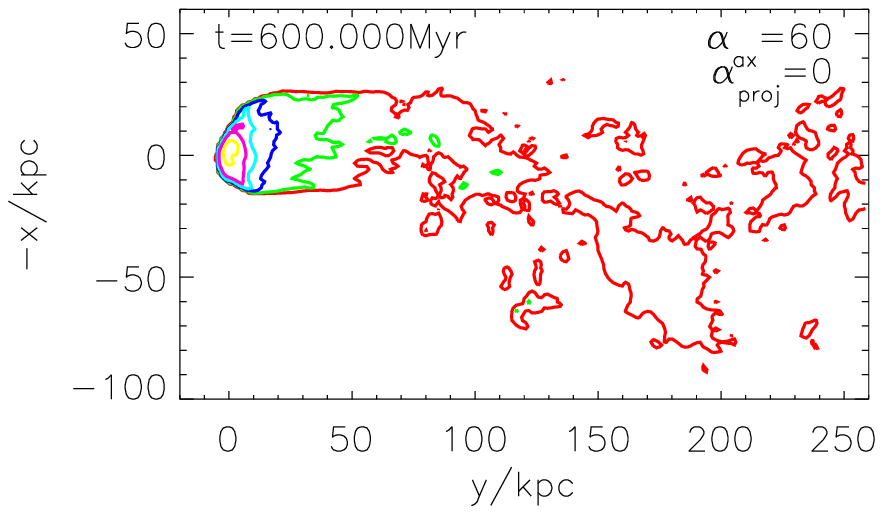}
}
\caption{Projected galactic gas density for medium ram-pressure, Mach number
  2.53, $i=30\degree$. Otherwise same as Fig.~\ref{fig:sdens_subsonic}.}
\label{fig:sdens_supersonic}
\end{figure}

\begin{figure}
\centering\resizebox{\hsize}{!}%
{\includegraphics{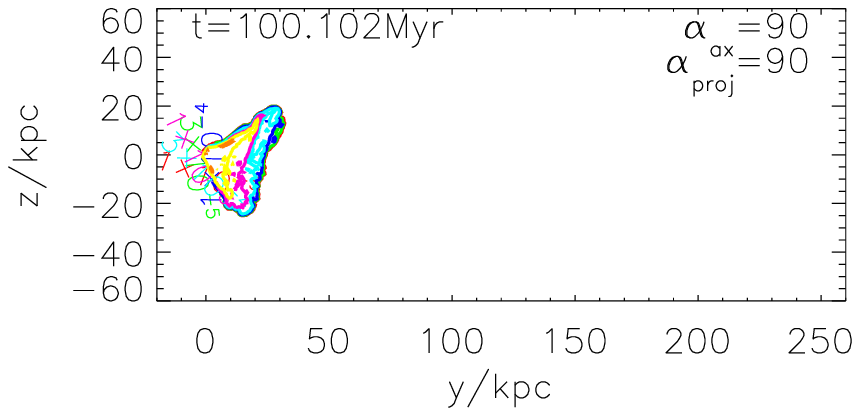}
\includegraphics{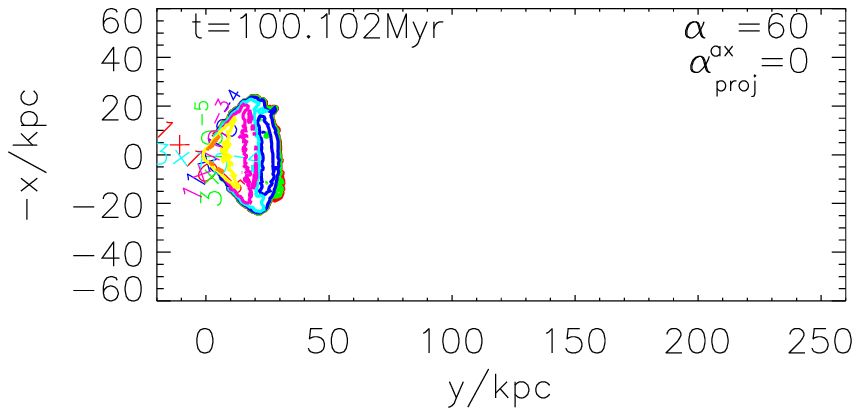}
}
\centering\resizebox{\hsize}{!}%
{\includegraphics{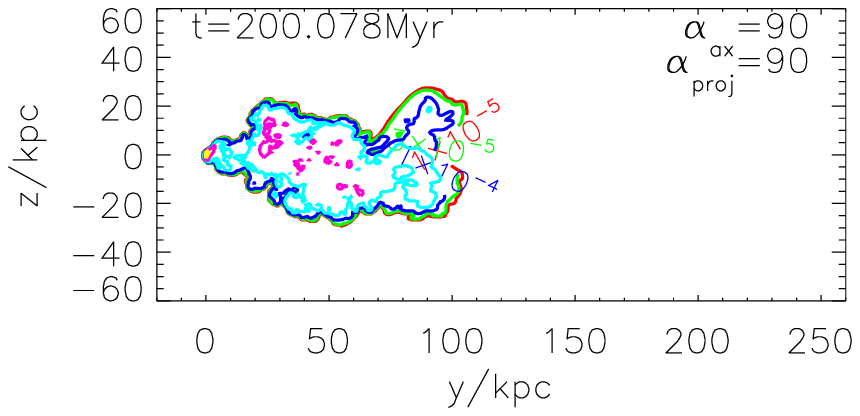}
\includegraphics{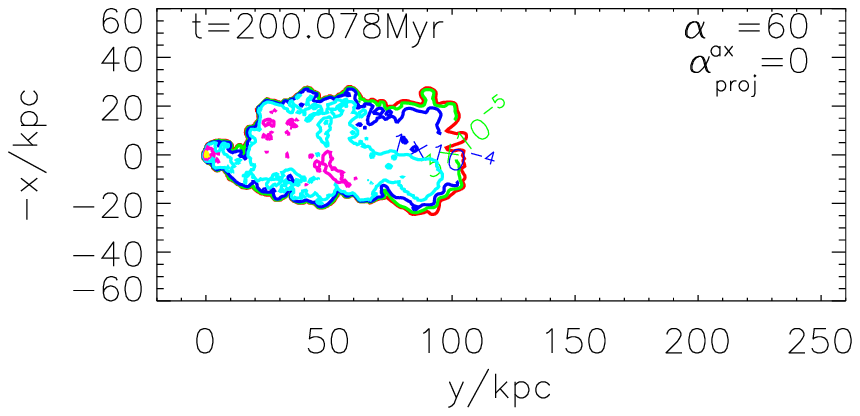}
}
\centering\resizebox{\hsize}{!}%
{\includegraphics{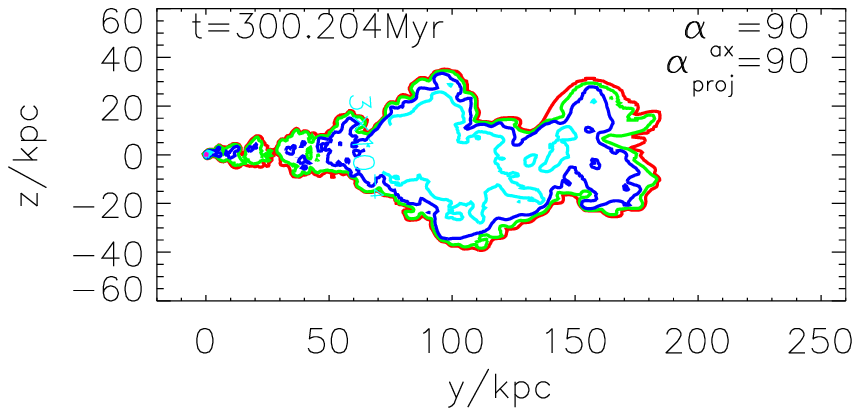}
\includegraphics{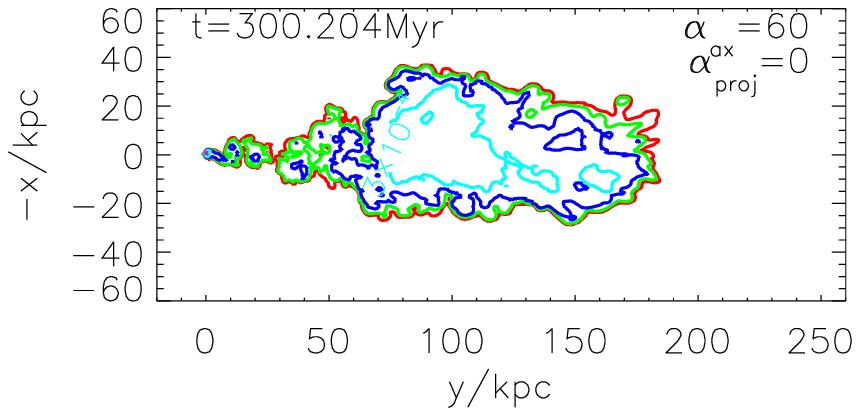}
}
\centering\resizebox{\hsize}{!}%
{\includegraphics{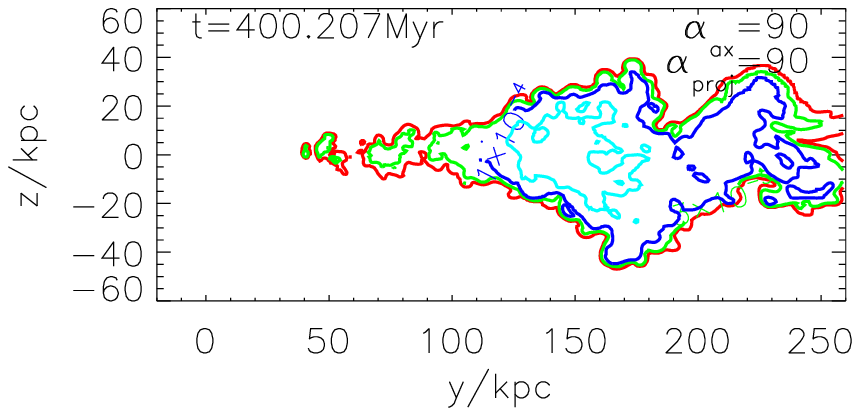}
\includegraphics{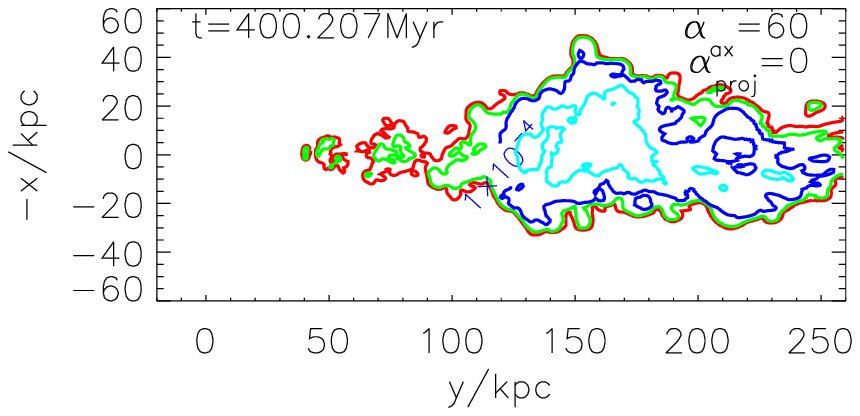}
}
\centering\resizebox{\hsize}{!}%
{\includegraphics{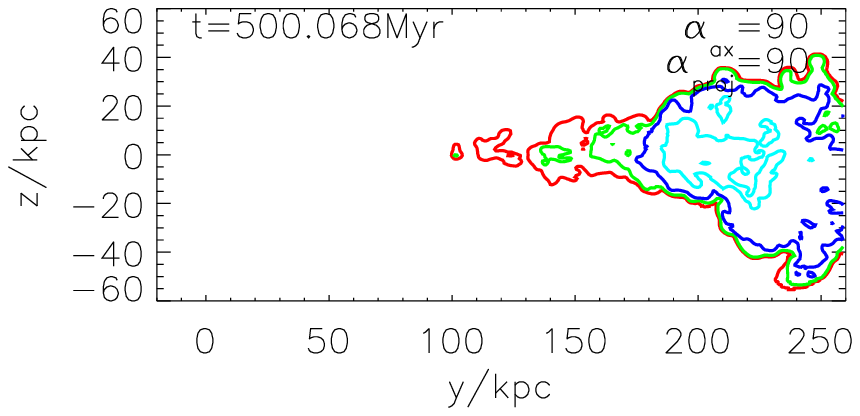}
\includegraphics{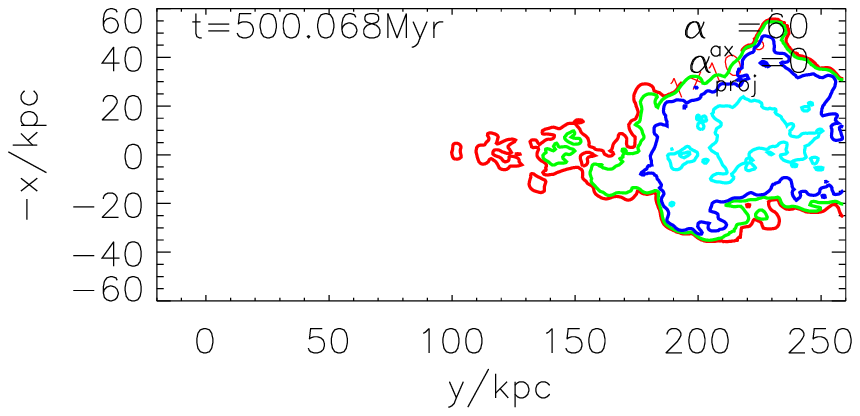}
}
\caption{Projected galactic gas density for strong ram-pressure, Mach number
  0.8, $i=30\degree$. Otherwise same as Fig.~\ref{fig:sdens_subsonic}. In the
  lower panels, the bulk of stripped gas can be seen to move away from the
  galaxy (the galactic centre is located at $(x,y,z)=(0,0,0)$ for all
  timesteps).}
\label{fig:sdens_subsonicP2}
\end{figure}

\begin{figure}
\centering\resizebox{\hsize}{!}%
{\includegraphics{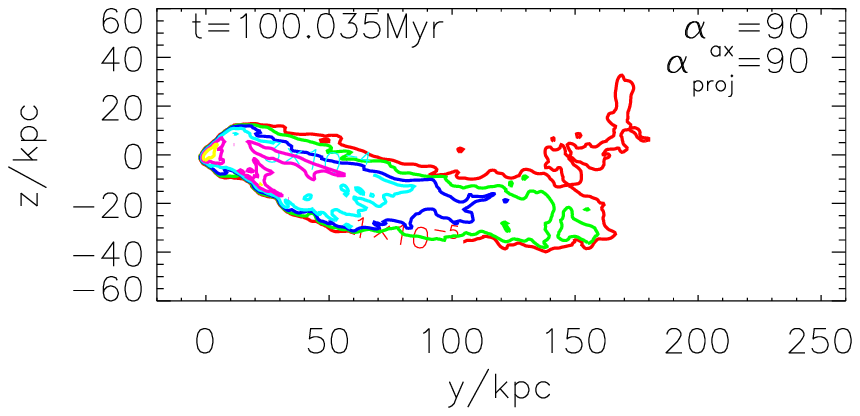}
\includegraphics{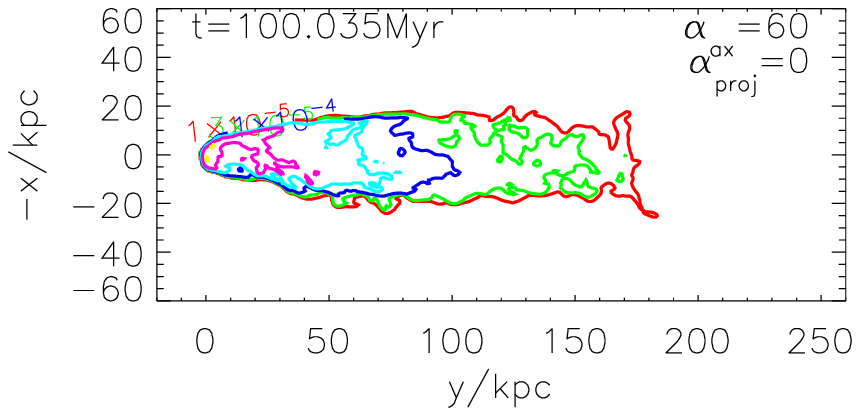}
}
\centering\resizebox{\hsize}{!}%
{\includegraphics{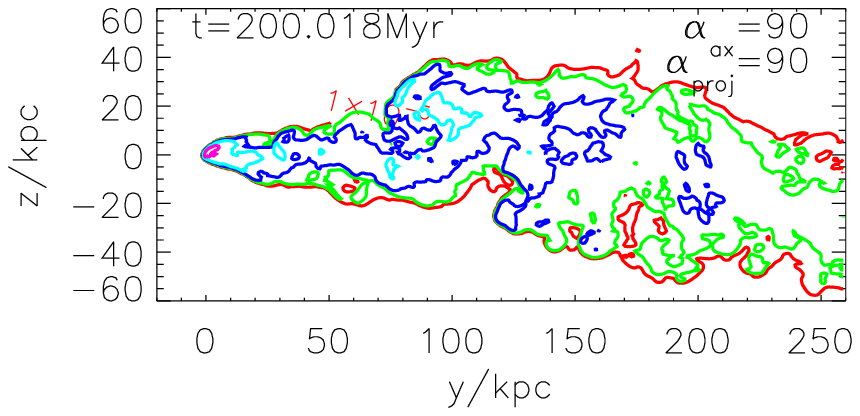}
\includegraphics{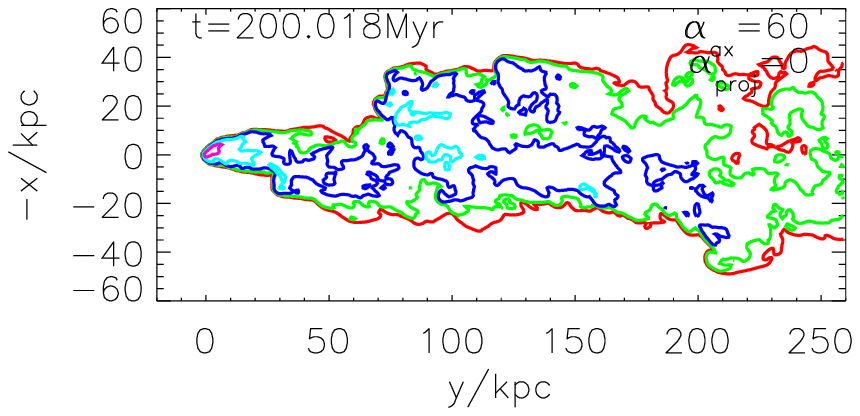}
}
\centering\resizebox{\hsize}{!}%
{\includegraphics{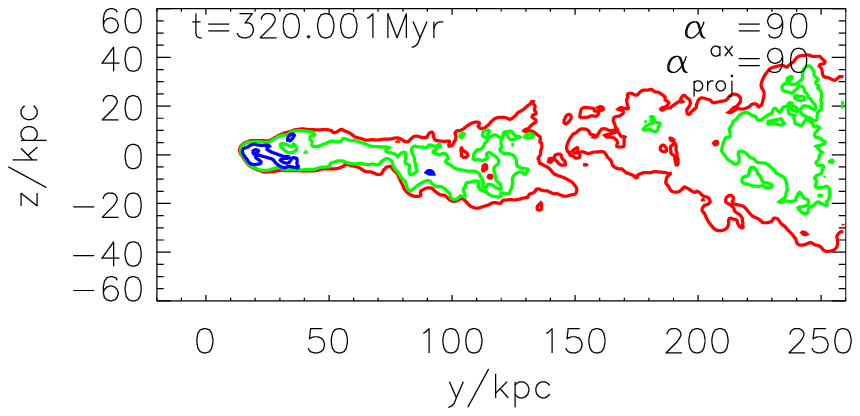}
\includegraphics{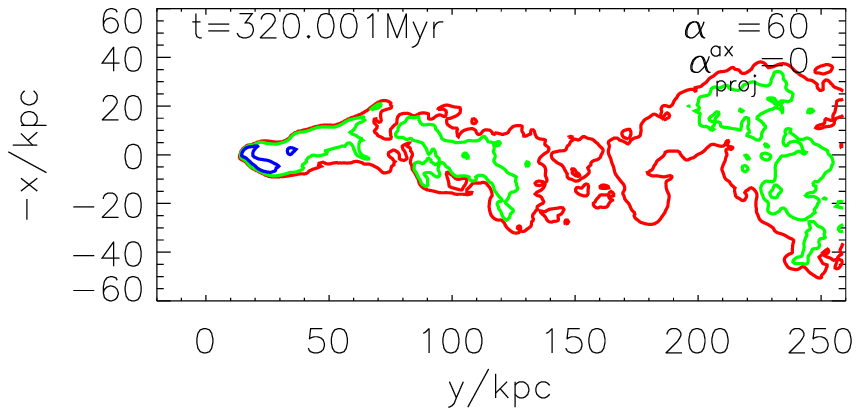}
}
\caption{Projected galactic gas density for strong ram-pressure, Mach number
2.53, $i=30\degree$. Otherwise same as Fig.~\ref{fig:sdens_subsonic}.}
\label{fig:sdens_supersonicP2}
\end{figure}

In this section, we show projected galactic gas densities along two
lines-of-sight for our simulation runs at different timesteps
(Figs.~\ref{fig:sdens_subsonic} to \ref{fig:sdens_supersonicP2}). The galactic
gas is identified by the mass scalar described in
Sec.~\ref{sec:code}. In all figures in this section, the left column shows
projections along the $x$-axis, while the right column shows projections along
the $z$-axis. These are the two directions perpendicular to the direction of
motion (which is along the $y$-axis). The time is denoted in the upper left
panel of each snapshot.  Each figure caption denotes the ram pressure, Mach
number and inclination angle of the run.
%
%
%
%
%
%
%

All projections of the medium ram-pressure cases show gas tails with column
number densities around $10^{19}\,\CM^{-2}$. The tails of the high ram-pressure cases have a column density roughly a factor of 10 higher.  Typical
tail widths are between 60 and $100\Kpc$.

In the cases with medium ram-pressure, the galaxy can retain a tail for several
$100\Myr$ and the tail can also evolve during this time. In the cases with
strong ram-pressure, the complete gas disc is stripped after a few $100\Myr$,
so that one cannot speak of a galactic tail any more. Instead, the bulk of
stripped gas moves downstream as a stretched gas cloud which is few $100\Kpc$
long. The fact that in the high ram-pressure cases much more gas is lost in a
shorter time leads to a distribution of more mass into a smaller volume and
thus higher column densities in the tails.

\subsection{Velocity structure in the tail} \label{sec:vel-structure}
\begin{figure}
\centering\resizebox{\hsize}{!}%
{
\includegraphics[width=0.24\textwidth]{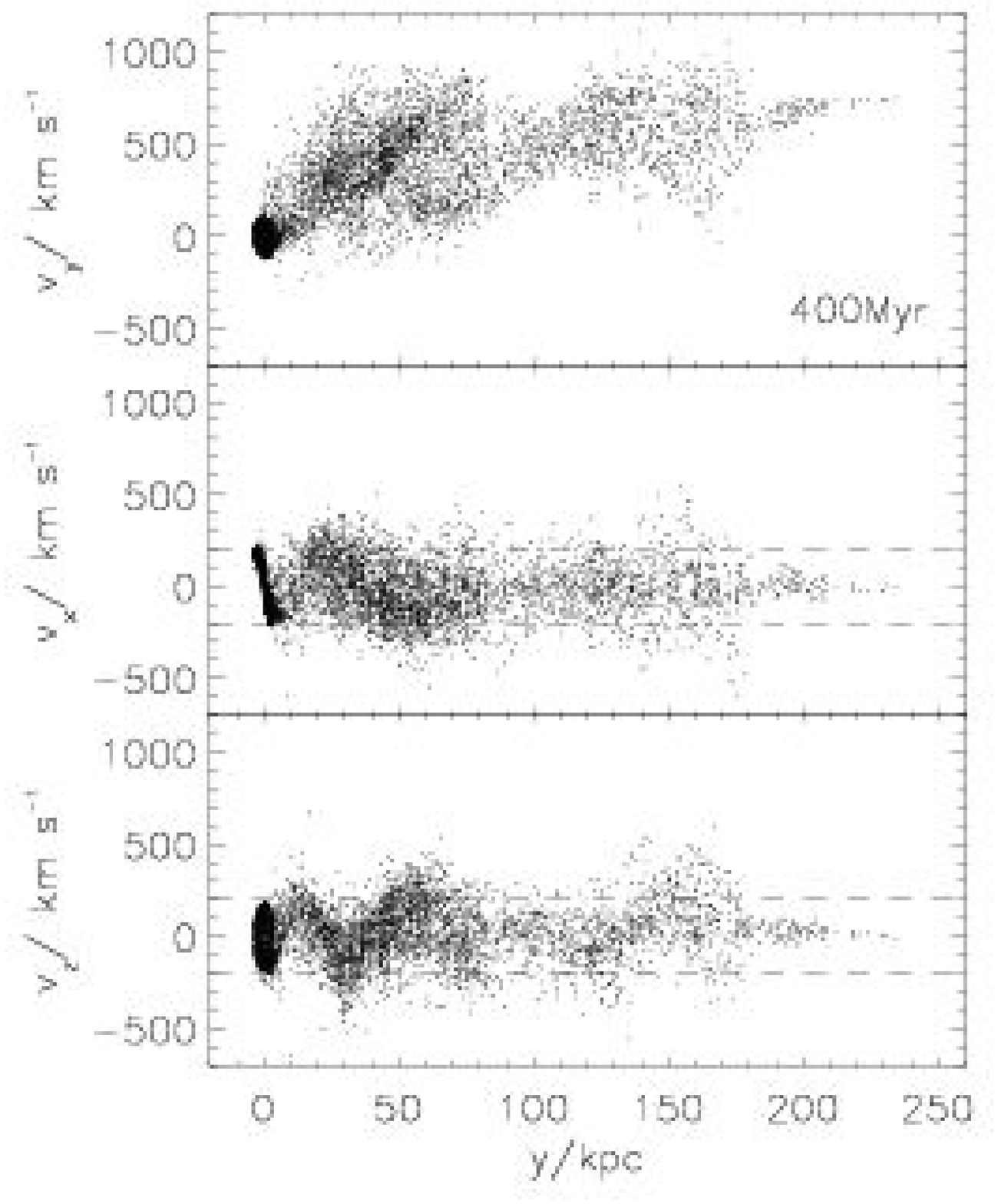}
\includegraphics[width=0.24\textwidth]{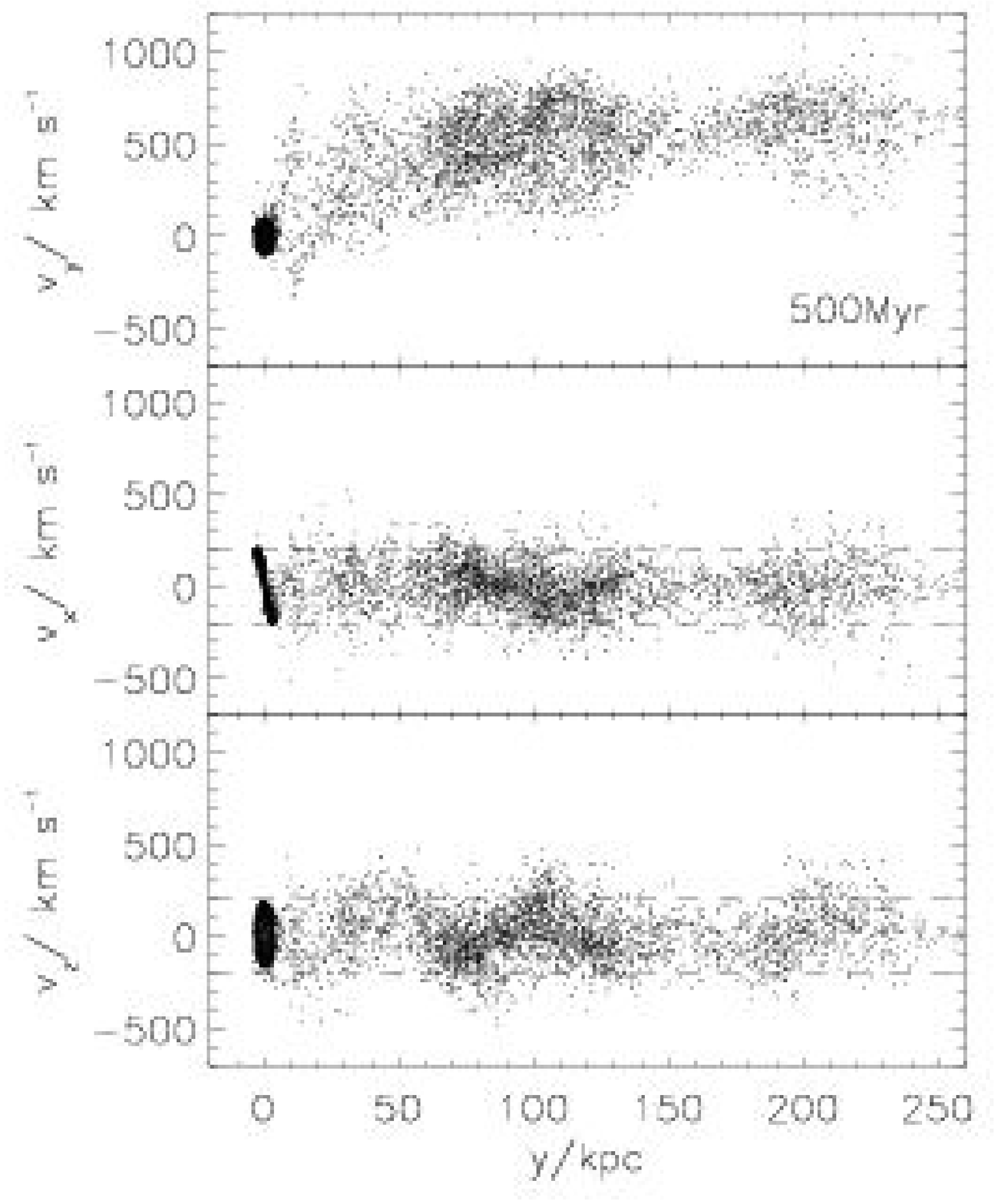}
}
\caption{Velocity components in the wake, for medium ram-pressure, Mach number
  0.8, $i=30\degree$. The top panels shows the velocity along wind direction
  ($y$-axis), the middle and bottom panels show the velocity components in the
two directions perpendicular to the wind direction ($x$- and
$z$-direction). Each column is for one timestep. The time is denoted in the
top panel of each column. The dashed vertical lines in the middle and bottom
panels denote the amplitude of galactic rotation.}
\label{fig:vel_subsonic}
\end{figure}

\begin{figure}
\centering\resizebox{\hsize}{!}%
{
\includegraphics[width=0.24\textwidth]{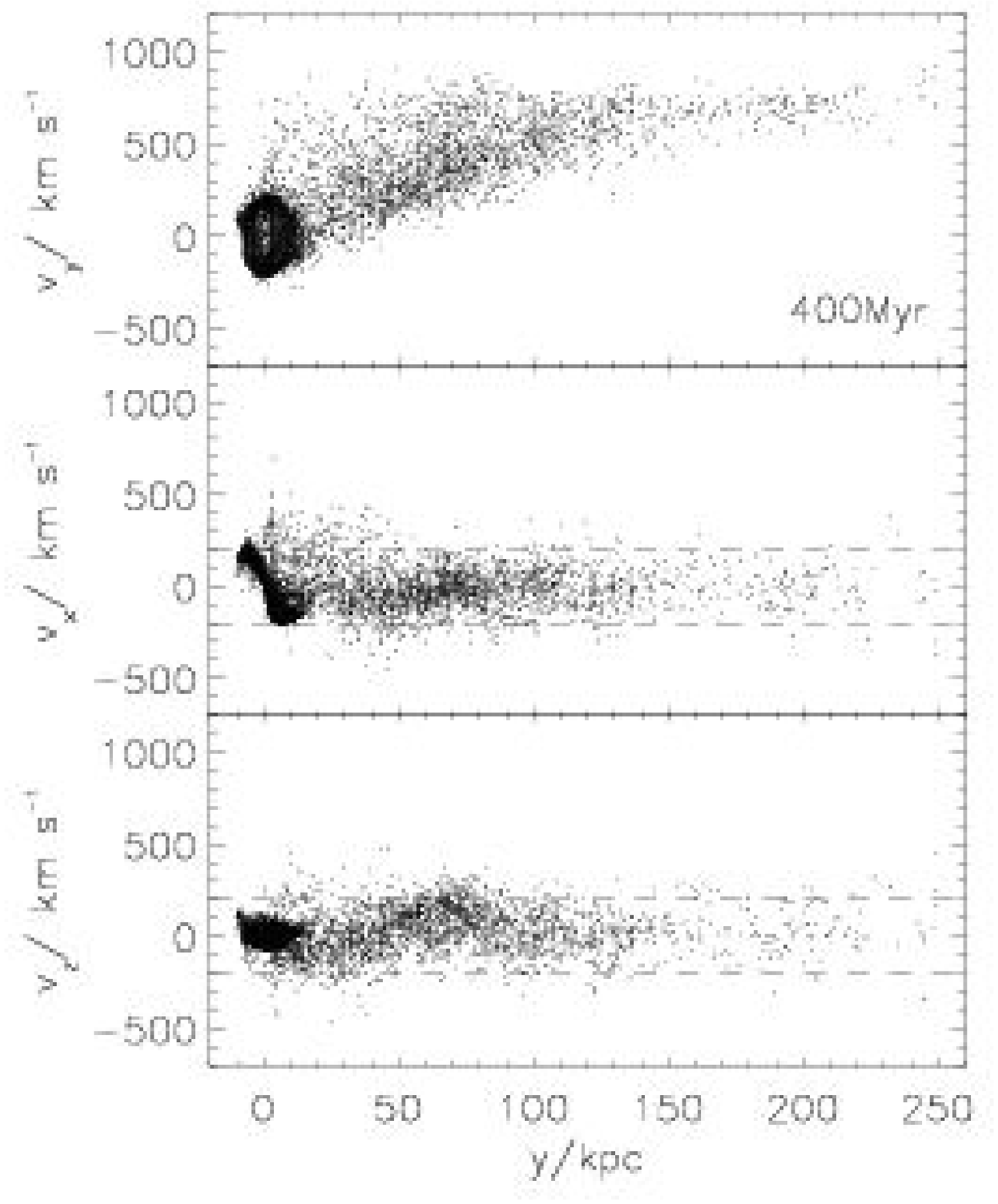}
\includegraphics[width=0.24\textwidth]{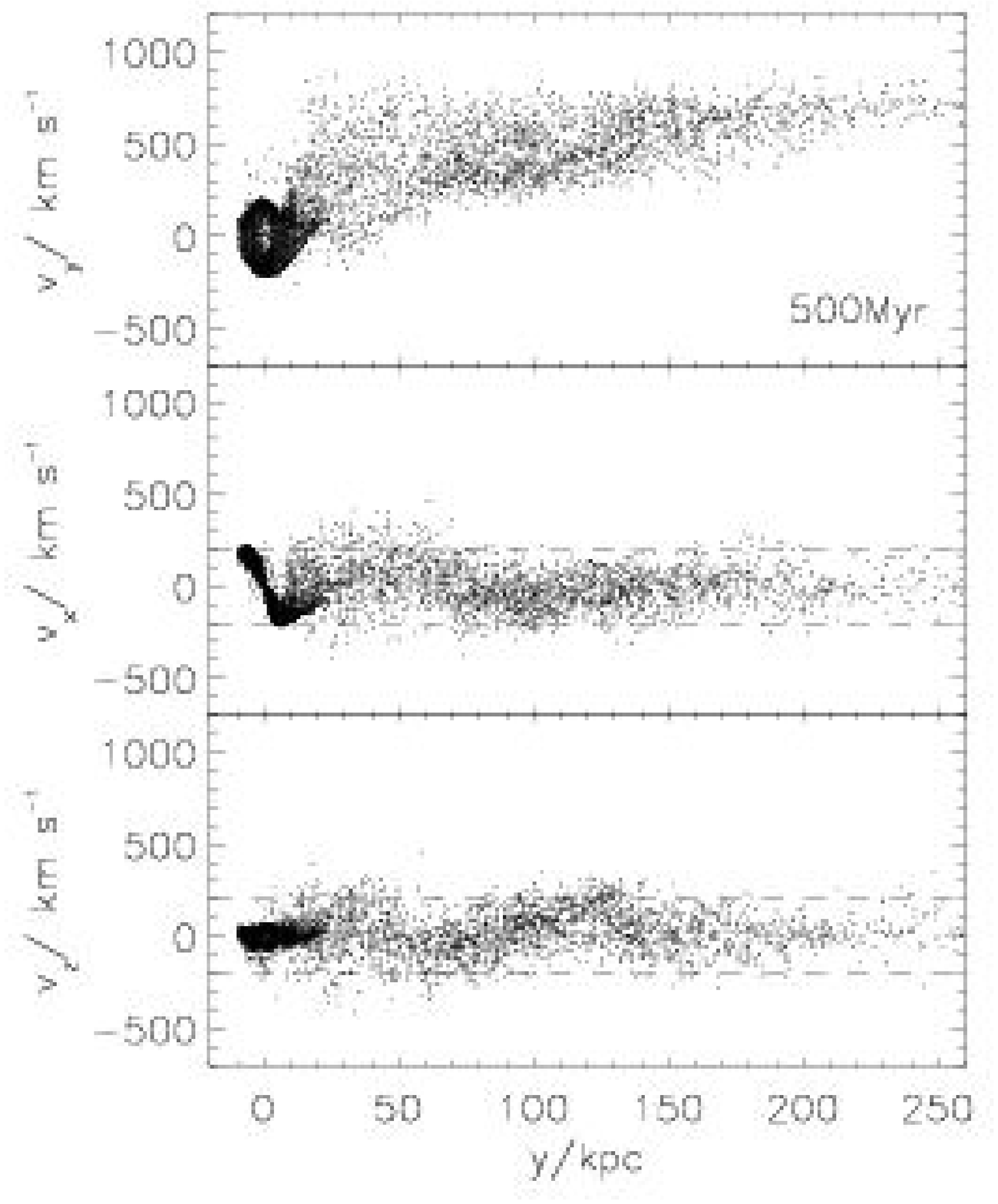}
}
\caption{Velocity components in the wake, for medium ram-pressure, Mach number
  0.8, $i=75\degree$. Otherwise same as Fig.~\ref{fig:vel_subsonic}.}
\label{fig:vel_subsonicI75}
\end{figure}

\begin{figure}
\centering\resizebox{\hsize}{!}%
{\includegraphics{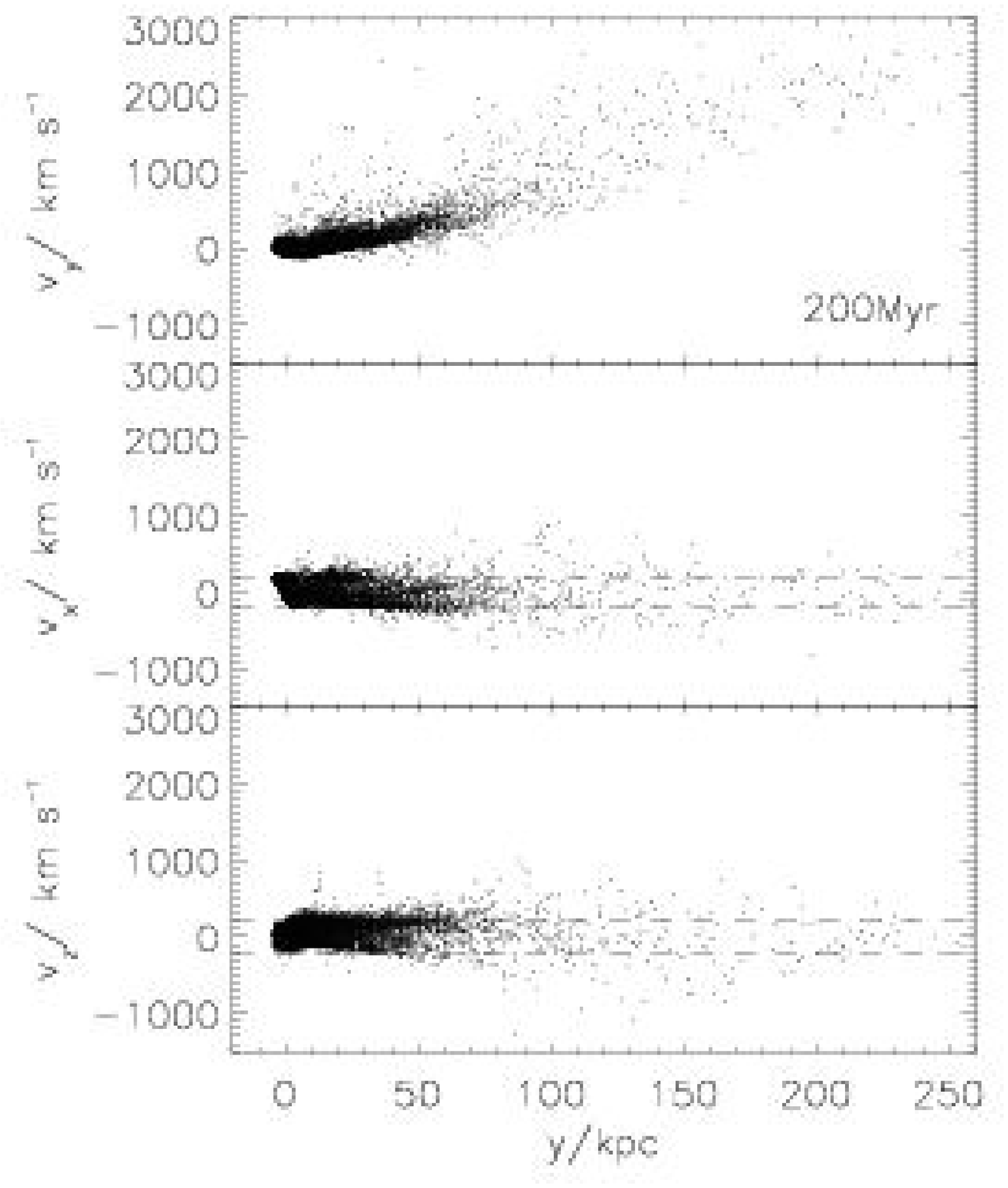}
\includegraphics{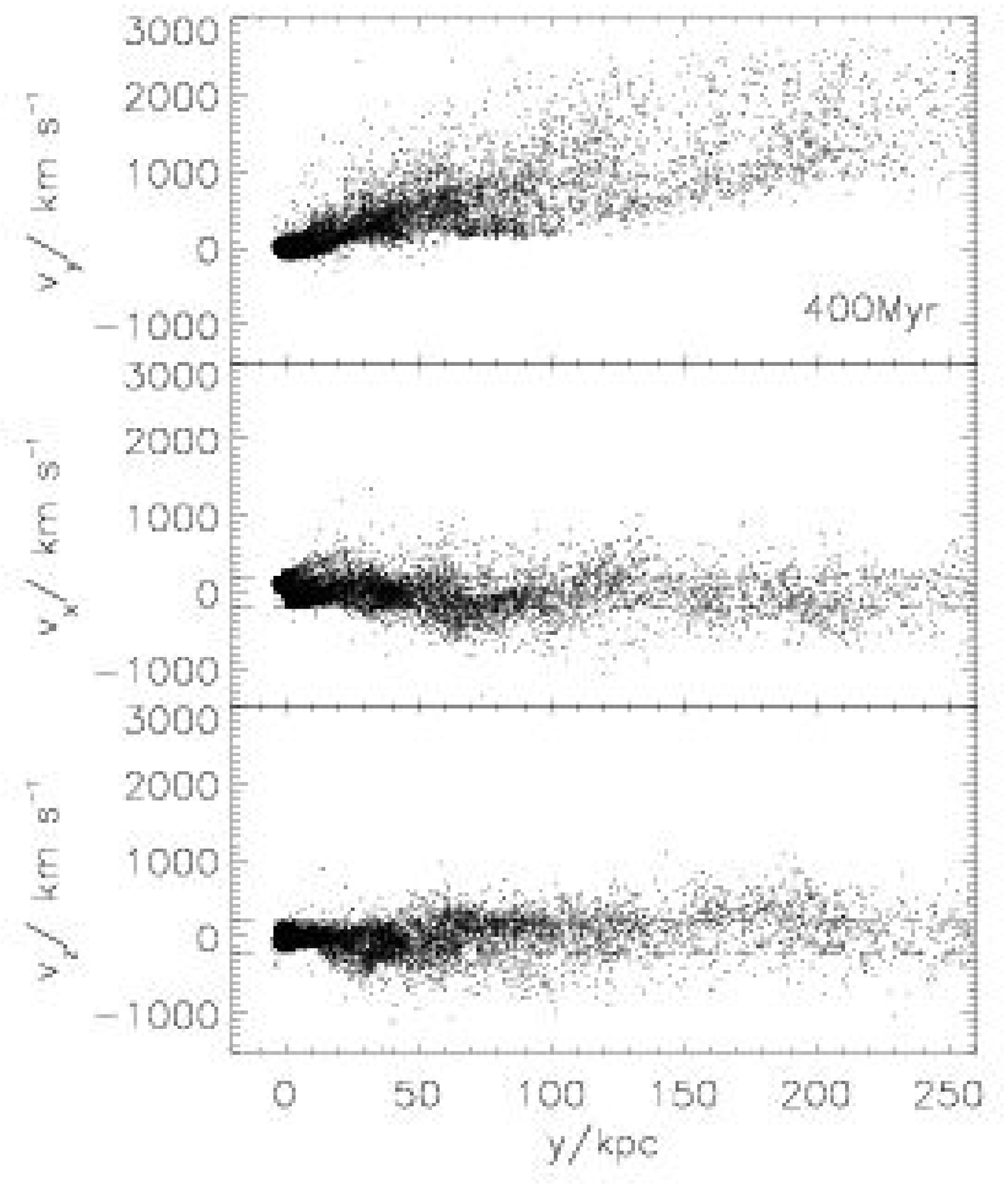}
}
\caption{Velocity components in the wake, for medium ram-pressure, Mach number
  2.53, $i=30\degree$. Otherwise same as Fig.~\ref{fig:vel_subsonic}.}
\label{fig:vel_supersonic}
\end{figure}

\begin{figure}
\centering\resizebox{\hsize}{!}%
{
\includegraphics[width=0.24\textwidth]{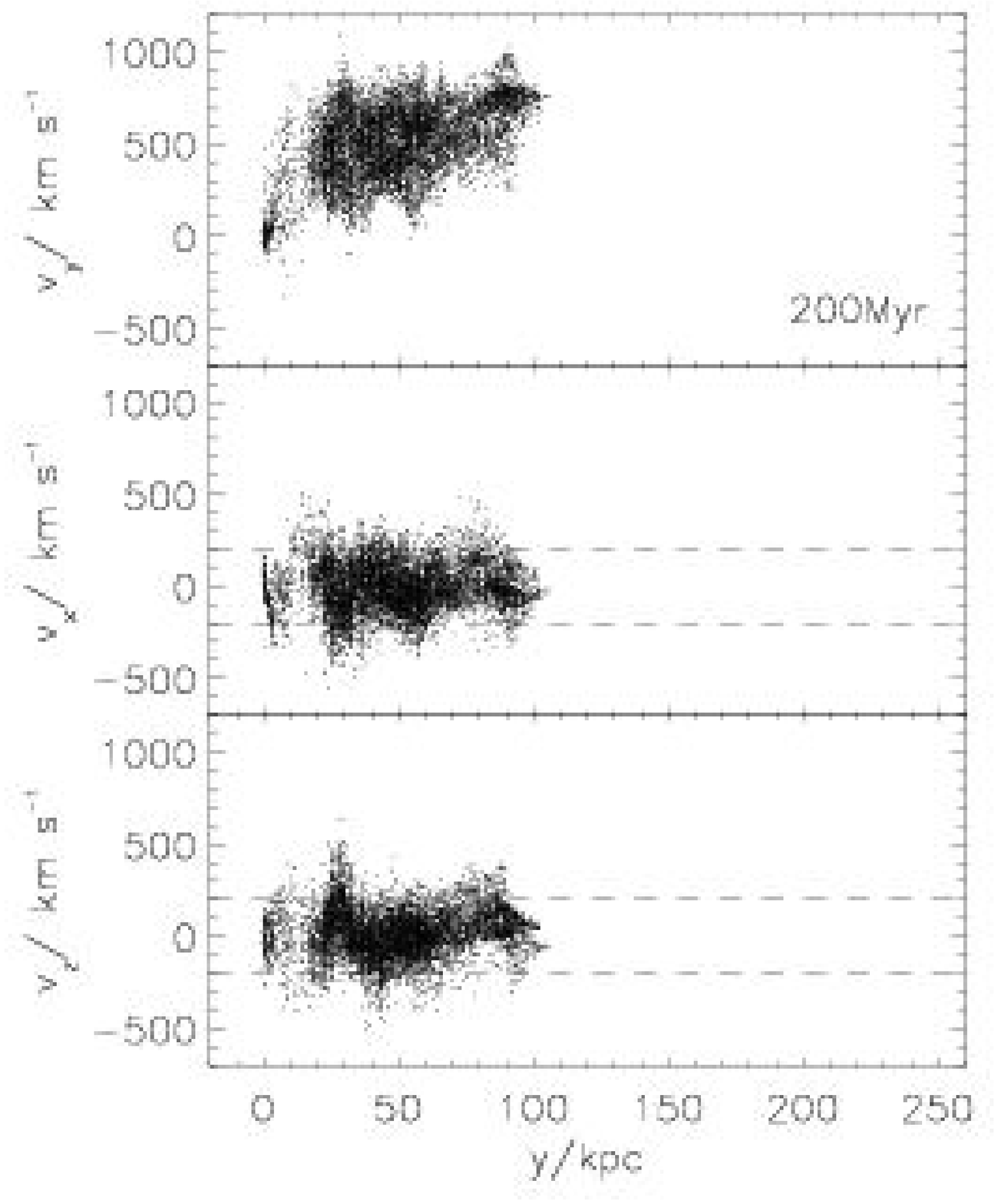}
\includegraphics[width=0.24\textwidth]{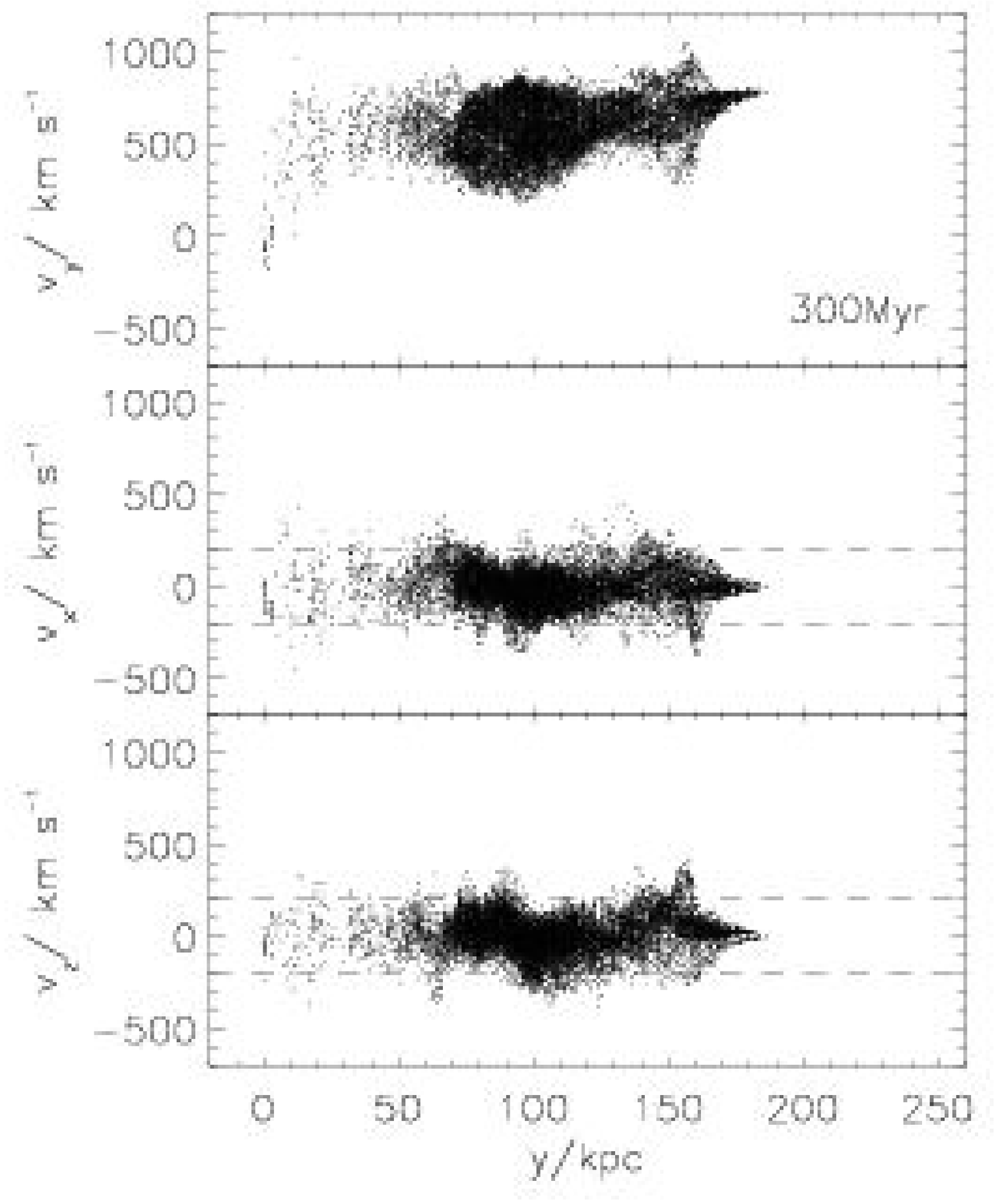}
}
\caption{Velocity components in the wake, for strong ram-pressure, Mach number
  0.8, $i=30\degree$. Otherwise same as Fig.~\ref{fig:vel_subsonic}.}
\label{fig:vel_subsonicP2}
\end{figure}

\begin{figure}
\centering\resizebox{\hsize}{!}%
{
\includegraphics[width=0.24\textwidth]{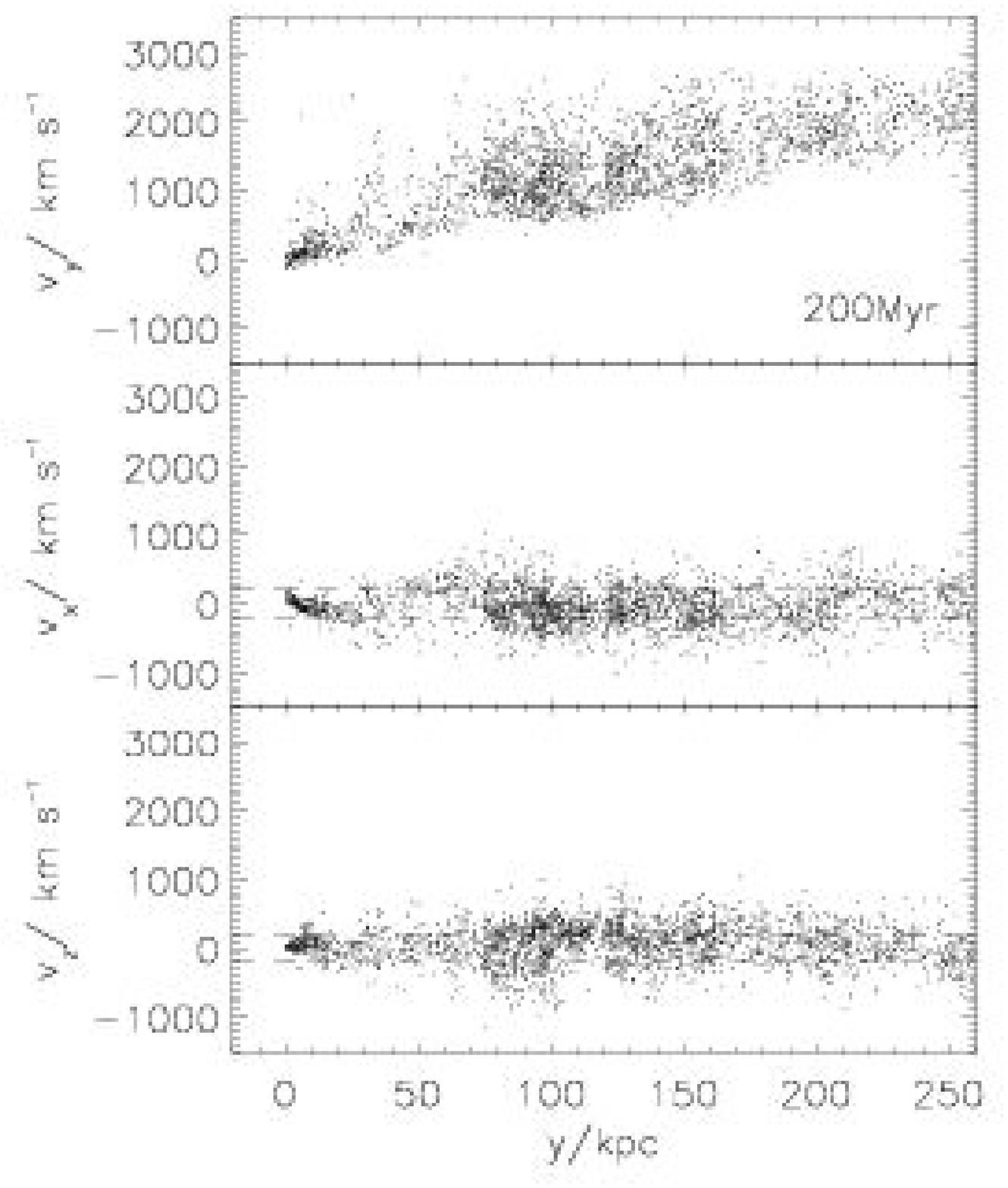}
\includegraphics[width=0.24\textwidth]{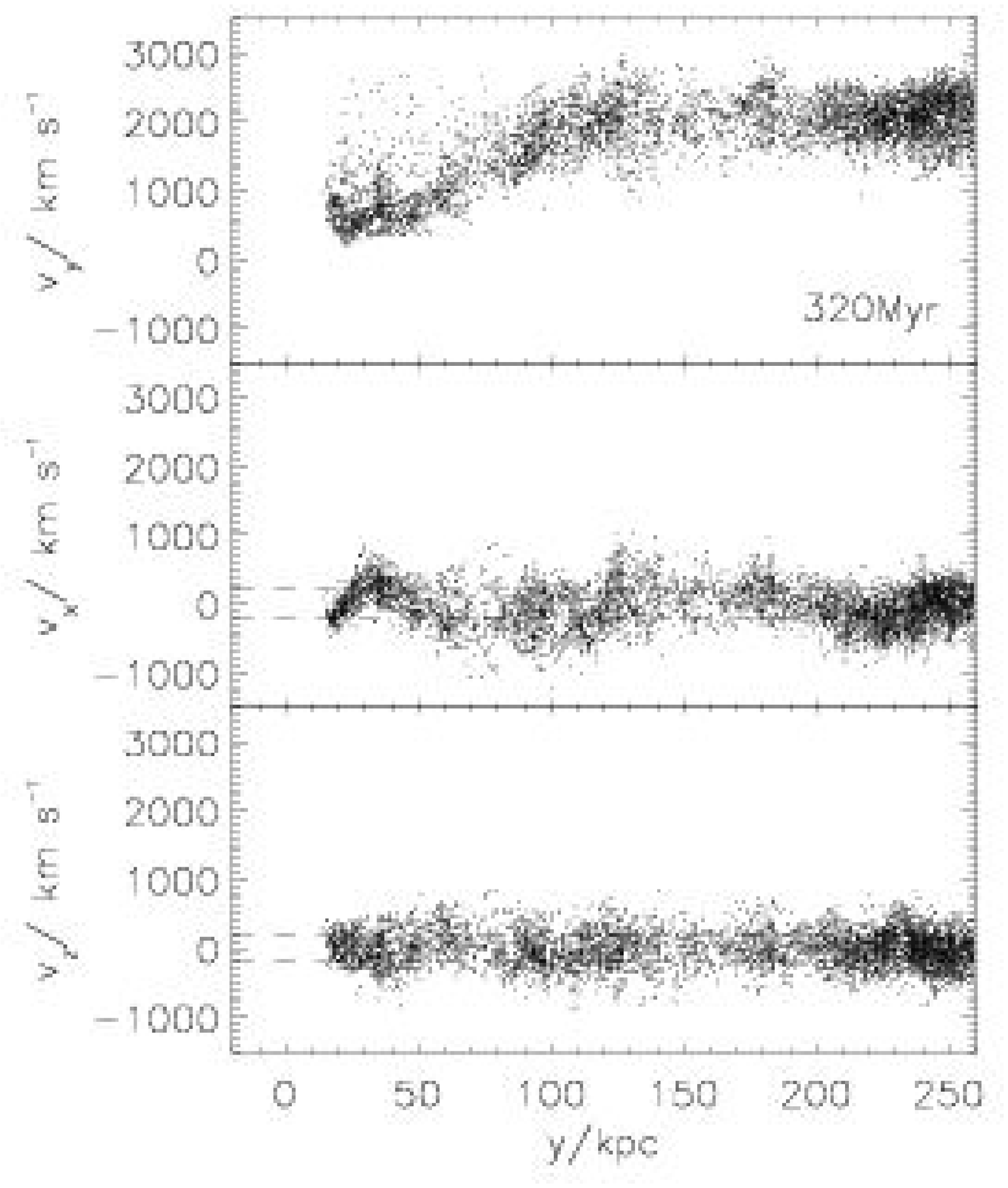}
}
\caption{Velocity components in the wake, for strong ram-pressure, Mach number
  2.53, $i=30\degree$. Otherwise same as Fig.~\ref{fig:vel_subsonic}.}
\label{fig:vel_supersonicP2}
\end{figure}

Figures~\ref{fig:vel_subsonic} to \ref{fig:vel_supersonicP2} show the
velocities of the stripped gas in the wake. For a random, ISM density weighted
subset of grid cells, the velocity components in $y$-,
$x$- and $z$-direction are plotted as a function of distance to the
galaxy. In our simulations, $v_y$ is the velocity component in wind direction,
whereas $v_x$ and $v_z$ are perpendicular to the wind direction. The caption of each figure denotes the ram pressure, Mach number and
inclination angle of the run.
%
%
%
%
%
%

The component $v_y$ reflects the acceleration of the stripped gas in the wind
direction. The components perpendicular to the wind direction cannot result
solely from galactic rotation. In the panels for the $v_x$- and
$v_z$-components, we have marked the amplitude of the galactic rotation
velocity $v\Rot=\pm 200\Kms$. If the stripped gas were pushed straight into
the wind direction, no velocities higher than this should occur in $v_x$ and
$v_z$. But in all cases velocities higher than the galactic rotation
occur. Especially in the $75\degree$ case, the galactic rotation induces only
a small $v_z$-component. Almost all of the observed $v_z$ must be due to the
ICM flow. The excess in $v_x$ and $v_z$ arises because the ICM is flowing
smoothly around the galaxy. Thus, already the ICM flow near the galaxy
contains velocity components in $x$- and $z$-direction. In the supersonic
case, the oblique bow shock also leads to non-zero $v_x$- and
$v_z$-components. The strongest increase of $v_x$ and $v_z$ is observed in the
supersonic case. Additional excess in $v_x$ and $v_z$ is introduced by
turbulence.

The velocity width in the tail also differs in subsonic and supersonic
cases. The overall width in the subsonic cases is $\sim 400$-$600\Kms$,
whereas in the supersonic cases it is nearly twice as much. The velocity width
is roughly independent of ram pressure.

Interestingly, in the subsonic cases, oscillations appear in $v_x$ and
especially in $v_z$. Their origin is discussed below.

\subsection{Tail structure}
The shape of the tail depends mainly on Mach number. Only in the
subsonic cases regular structures similar to a von Karman vortex street can be
produced (see Figs.~\ref{fig:sdens_subsonic} and
\ref{fig:sdens_subsonicI75}). The formation of a long regular tail also takes
several $100\Myr$. Therefore, less regular structure is produced in the strong
ram-pressure run because the gas disc is already stripped completely after a
few $100\Myr$. Moreover, in Fig.~\ref{fig:sdens_subsonic}, an
oscillating tail appears in the projection along the $x$-axis, but not along
the $z$-axis. If we
neglect the disc rotation for the moment, then the galaxy
appears asymmetrical due to its inclination only in $z$-direction, but not
in $x$-direction. Therefore, the ``von Karman oscillations'' are more
prominent in $z$-direction, and hence can be seen best if the line-of-sight is
the $x$-axis. Also the oscillations observed in the velocity components
perpendicular to the wind direction $v_x$ and $v_z$ (see
Sect.~\ref{sec:vel-structure}) are connected to the ``von Karman
oscillations''.

The supersonic cases show an irregular tail structure.

\subsection{Flaring of the wake}
In addition to the von-Karman oscillations, in the subsonic cases the
tail widens systematically although this effect is less prominent in
the high ram-pressure case because the gas disc is already stripped
completely at early times. The velocity components perpendicular to
the wind direction are similar for, both, the medium and high
ram-pressure case. Therefore, the tail is slightly narrower in the
stronger ram-pressure case because here the stripped gas is
accelerated faster.

In the subsonic, medium ram-pressure cases, the widening proceeds
roughly linearly on the length scale of our simulation box. The tail
width ranges from a few $10\Kpc$ near the galaxy to $150\Kpc$ at the
end of the simulation box (distance of $260\Kpc$ behind the
galaxy). The diagonal lines in Figs.~\ref{fig:sdens_subsonic} and
\ref{fig:sdens_subsonicI75} envelop the tail. They were fitted by
eye. In each figure, they are identical for all panels, i.e. they are
independent of time and projection direction.

The difference in the angles of the envelopes between the two
inclinations is very small. However, for the more highly inclined
galaxy, the enveloping lines are closer together. For higher
inclinations, the width of the tail at a given distance to the galaxy
is smaller. In both inclinations, the upper envelope has an angle of
$13\degree$ with respect to the $y$-axis. The angles between the lower
envelope and the $y$-axis are $9\degree$ and $10.5\degree$ for the
inclinations of $75\degree$ and $30\degree$, respectively. For the
projections along the $z$-axis (right column) the different angles
between the lower and upper line could be explained by the disc's
rotation. In these panels, the disc is rotating
counter-clockwise. Therefore, at the bottom side of the panel, the
disc is rotating in wind direction, whereas on the upper side of the
panel the disc is rotating antiparallel to the wind direction. This
leads to a slightly stronger stripping at the bottom side and a
slightly stronger stretching of the bottom side of the tail.  For the
projection along the $x$-axis, the reason must be a different one
because here the rotation does not induce a similar asymmetry. Here,
the inclination makes the galaxy asymmetrical for the ICM
flow. Moreover, the stripping process itself proceeds asymmetrically
for inclined galaxies (see
\citealt{roediger06}). However, it is surprising that the asymmetry of the
tail is similar for both inclinations and both projection directions.

The fact that the widening of the tail proceeds in a very similar fashion
independent of galactic inclination suggests that the dynamics of the widening
are independent of the galaxy but intrinsic to the ICM flow. This also
excludes the widening due to the rotation of the disc gas as presumed in the
introduction as a major process. 

In the supersonic, medium ram-pressure case the tail widens, too,
despite its irregular behaviour. The least widening is observed in the
supersonic high ram-pressure case. However, the velocity components
perpendicular to the wind direction are high in the supersonic case,
so that some widening could be observed if the tail is traced to even
larger distances behind the galaxy.

\section{Discussion}
%
\subsection{Observable galactic tails}
%
\subsubsection{General}
Given that the stripped gas remains in the HI temperature range, our
simulations suggest that deep HI observations could observe long galactic
tails. For galaxies that experience a strong ram pressure and lose nearly
their complete gas disc in a few $100\Myr$, the tails could be detected with
less sensitive observations. In such cases, the tail can even separate from
the source galaxy (see bottom panels of Fig.~\ref{fig:sdens_subsonicP2}) and
appear as a large isolated gas cloud. In addition to the spatial separation,
the ``tail'' gas cloud will also be separated from its source galaxy in
velocity space. This makes it very difficult to recognise the original
connection between the gas cloud and the source galaxy. An HI survey of the
Virgo cluster (\citealt{davies04a}) has indeed found two candidates of HI
clouds without optical counterparts.

Our simulations predict tail widths in the range of 60 to $100\Kpc$ at
\emph{non-projected} distances of $\sim 100\Kpc$ behind the galaxy. If a
galaxy moves subsonically, if it is exposed to a medium ram pressure
and seen along a favourable line-of-sight, even regular structures
similar to von-Karman oscillations could be detected. However, some
additional processes could modify the appearance of the tails:

\begin{itemize}
\item For galaxies passing through cluster centres, the ICM wind will be
  variable. The decrease of ram pressure after the peak ram pressure and
  back-falling material could change, both, the tail structure and the
  structure of the velocity component in wind direction. The velocity
  components perpendicular to the wind direction should not be sensitive to
  this process. However, in observations the components into and perpendicular
  to wind direction may be difficult to disentangle.
\item We neglected cooling and thermal conduction. These processes are
  important for the question whether the stripped gas stays observable as
  HI. If a tail is observed during heating or cooling, the structure observed
  in HI may be completely decoupled from the overall structure of all stripped
  gas because some parts drop out of the relevant temperature range. However,
  the observation of \citet{oosterloo05} suggests that at least a part of the
  stripped gas can survive as HI for a few $100\Myr$. More detailed studies
  are needed to investigate the role of thermal conduction.
\item We also neglected viscosity in our simulations. Due to numerical
  viscosity, our simulations match the case of moderate to high Reynolds
  numbers. Magnetic fields in the ICM may suppress viscosity significantly,
  which could lead to higher turbulent velocities and greater widening of the
  tail. On the other hand, a substantial amount of viscosity
  would smooth the turbulent structures that develop in our simulations, and
  galactic tails may be narrower than we predict.
\item Local ICM motions as found in many cluster simulations
  (e.g.~\citealt{faltenbacher05,takizawa05a}) could deform and bend the
  galactic tails.
\end{itemize}

\subsubsection{The case of NGC 4388}
So far, the only example of a long tail of cool ram pressure stripped gas is
the case of NGC~4388 in the Virgo Cluster (\citealt{oosterloo05}). This
galaxy's tail has a length of $\sim 120\Kpc$ in the plane of the sky, the tail
width flares from about $15\Kpc$ near the galaxy to about $40\Kpc$ at a
projected distance of $100\Kpc$ behind the galaxy. The ridge of maximum flux
density suggests a slight S-shape of the tail. 

In order to disentangle the true properties of the wake, the velocity
component in the plane of the sky would have to be known. The length of the
tail indicates that this component is substantial. The radial velocity of 
NGC~4388 with respect to the cluster mean is $\sim 1400\Kms$ (see
\citealt{vollmer03a} and references therein). Even if this galaxy had no
velocity component in the plane of the sky, its radial velocity suggests
already a slight supersonic motion through the ICM, which would be enough to
qualify this galaxy's velocity as typical for cluster galaxies. Consequently,
a substantial velocity component in the plane of the sky as indicated by the
long tail suggests that NGC~4388's true velocity with respect to the cluster
mean is rather high. Thus, it is very likely that this galaxy is moving
supersonically through the ICM. Therefore, our simulations suggest that the
implied S-shape in the tail is not due to regular von-Karman oscillations. The
observed S-shape could also be coincidental and not part of a regular pattern.
In our supersonic simulations situations like this occur, the supersonic tails
are far from being perfectly straight. NGC~4388's tail shape may also be caused by other
processes discussed above. 

Even if the velocity component in the plane of the sky is comparable to the
radial component (in the cluster rest frame), this implies that the true
length of the tail is at least $1.4$ times as long as the projected one. If
the velocity component in the plane of the sky is only $\sim 500\Kms$, the
true tail is about 3 times as long as the projected one. A reasonable
assumption for the ICM density at NGC~4388's position is a few
$10^{-4}\ccm$. With an assumed velocity of $\sim 1800\Kms$ it is likely that
this galaxy experienced a ram pressure in the range we studied here. In our
simulations, we can reproduce such very long tails. The details of the tail
length may also depend on the exact stripping history, i.e. the time
dependence of the ram pressure during the cluster passage. Thus, comparisons
with simulations of realistic cluster passages that also trace the evolution
of the galactic tail may be used to infer the galaxy's proper velocity.

The width of the tail does not depend on the projection direction. Thus, the
tail of NGC~4388 is narrower than the tails in our simulations. This
difference may be caused by reasons discussed above. E.g.~if the ICM is more
viscous than in our simulations, the tail will be narrower. Alternatively,
some parts of the tail may already have dropped out of the relevant
temperature range in order to be observable in HI. Comparisons with
simulations including these effects could give important insights on the
degree of viscosity and thermal conduction in the ICM.

\citet{oosterloo05} have also measured the radial velocity in the tail of NGC
4388. Close to the galaxy, the radial velocity agrees with the galaxy's radial
velocity. Then it decreases along the tail by about $550\Kms$. This means that
the stripped gas at the end of the tail has reached only about 40\% of the
galactic radial velocity with respect to the cluster mean. However, it takes
some time until the stripped gas is decelerated sufficiently to fall behind
the galaxy. In this point, our simulations agree with the observations. Also
in our supersonic simulations, the stripped gas does not reach the full ICM
wind speed inside our simulation box, and gas at true distances of $100\Kpc$
behind the galaxy has reached only about 50\% of $v\ICM$.  Again, details of
the velocity structure depend on the exact stripping history.

In addition to the velocity decrease, \citet{oosterloo05} have measured the
velocity width in the tail. They derive values of 100 to $200\Kms$. This is
surprisingly low as one would expect the stripped gas to preserve the rotation
it had inside the galaxy, i.e. the width should be at least twice the rotation
velocity. Our simulations show at least this width, and the supersonic
simulations even show a higher width. This is in accord with the simulations
of \citet{vollmer03a} that produce velocity widths of $\sim 400\Kms$. The
reason for this discrepancy might be that \citet{oosterloo05} have measured
the velocity only along the ridge of highest flux density. Thereby, they may
have incidentally selected a special fraction of the tail, like gas that
originated from a certain part of the galaxy only.

\subsubsection{Other examples}
To our knowledge, no other long galactic tails have been observed in HI. Some
observations exist in other wavelengths. E.g.~for CGCG~97-073 and 97-079 in
the cluster A1367 $H\alpha$ and radio continuum tails have been detected (see
\citealt{gavazzi95, gavazzi01}). Both galaxies show tails of $\sim 75\Kpc$
length. The H$\alpha$ tails are rather thin and straight and show little
substructure. The radio continuum tails show a some widening. It will be
interesting to observe these galaxies and others in deep HI for further
analysis.

\subsection{Metal distribution in the ICM due to ram pressure stripping}
Gas stripped from galaxies is an important source of metals for the ICM
(\citealt{schindler05,domainko05}). Therefore, the evolution of galactic tails
plays an important role in the distribution of metals in the ICM. In our
simulations, we find that the width of the tails behaves like $w/d\leq 0.4$
(see e.g.~Fig.~\ref{fig:sdens_subsonic}), where $w$ is the tail width and $d$
the distance from the galaxy. However, our simulations trace only the first
$260\Kpc$ behind the galaxy and the width of the wake is not going to increase
linearly with the distance from the galaxy. In a simple model, we can model
the turbulent widening of the wake as a diffusion process with a diffusion
constant given by $D\approx \frac{1}{3} \sigma l$, where $\sigma$ is a typical
turbulence velocity and $l$ a typical size of turbulence eddies. Our
simulations suggest velocity dispersions of $\sigma \sim 250\Kms$ for galactic
velocities of $\sim 800\Kms$ to $\sigma \sim 500\Kms$ for galactic velocities
of $\sim 2500\Kms$. The typical eddy size, $l$, in our simulations is $\sim
50\Kpc$. Thus, the time to reach a certain width $w$ is given by
\begin{eqnarray}
t &\sim& \frac{1}{4}\frac{w^2}{D} =\frac{3}{4}\frac{w^2}{\sigma l}\nonumber\\ 
&=& 600\Myr \left(\frac{w}{100\Kpc}\right)^2 \left(\frac{\sigma}{250\Kms}\right)^{-1} \left(\frac{l}{50\Kpc}\right)^{-1}.
\end{eqnarray}
During this time, the tail length increases by $\sim v\ICM t$. Consequently,
the tail length $d$ and its width relate as
\begin{eqnarray}
w &\sim& \sqrt{\frac{4 d D}{v\ICM}} = \sqrt{\frac{4d\sigma
    l}{3v\ICM}}\nonumber \\
&=& 130\Kpc \left(\frac{d}{1\Mpc}\right)^{1/2}
\left(\frac{v\ICM}{1000\Kms}\right)^{-1/2}\nonumber\\
&&\left(\frac{\sigma}{250\Kms}\right)^{1/2} \left(\frac{l}{50\Kpc}\right)^{1/2}
\end{eqnarray}

This estimate neglects that the turbulence decays as the gas streams
away from the galaxy. The decay of turbulence makes, both, $\sigma$
and $l$ smaller with increasing distance to the galaxy and therefore,
the above estimate gives an upper limit on $w$.

These estimates could be included in future simulations of metal
enrichment of the ICM by ram pressure stripping. Some of these
simulations inject metals provided certain analytical criteria for
ram-pressure stripping are fulfilled. These simulations could be
refined by imparting the stripped material with a velocity
dispersion to produce more realistic tails.



\subsection{Heating of the ICM by turbulence in the wakes}
The ongoing search for a heat source in galaxy clusters that prevents
catastrophic cooling in the cool cores makes it interesting to
estimate the extent to which the motion of galaxies can heat the ICM.

The kinetic energy of a galaxy can heat its ambient medium via three
mechanisms: (i) the flow around the galaxy can lead to turbulence in the wake
of the galaxy as seen in our simulations, (ii) dynamical friction accelerates
the surrounding gas (\citealt{ostriker99,kim05}), and (iii) in case of
supersonic motion, the ambient gas is heated by a bow shock. For the first two
mechanisms the induced motions need to be converted to heat by some
dissipative process.

As the galaxy moves through the ICM, it produces a turbulent wake. The
energy loss of the galaxy may be estimated by
\begin{equation}
  \left.
  \frac{{\rm d}E_{\rm gal}}{{\rm d} t}
  \right|_{\rm ram}
  =
  -
  p_{\rm ram} A_{\rm proj} v_{\rm ICM}
  ,
\end{equation}
where $A$ is the projected surface area of the galaxy. Using
conservative values, i.e. $p_{\rm ram} = 6.4 \times 10^{-11} \;{\rm
erg \, cm^{-3}}$, $v_{\rm ICM} = 1000 \:{\rm km \, s^{-1}}$ and $A = 4
\pi ( 4 \:{\rm kpc} )^2$, we obtain a loss rate $\sim 10^{42}\:{\rm
erg \, s^{-1}}$. This energy can be found as turbulent energy in the
ICM. For about 100 galaxies within a cluster, this heating rate is
comparable to the cooling rate by radiative losses.

Also, this heating rate is consistent with what we find in our
simulations. A rough inspection of the velocity distribution, see
Fig.~\ref{fig:denscut_supersonic} and \ref{fig:vel_supersonicP2},
shows that the ICM is stirred within a tube of diameter $\sim 40 \:
{\rm kpc}$ with a velocity dispersion of $\sigma_x \sim 400 \: {\rm km
\, s^{-1}}$. The ICM has an average gas density of $10^{-28} \: {\rm g
\, cm^{-3}}$ and the galaxy moves with $1000 \:{\rm km \, s^{-1}}$
through the ICM. This yields a rate of kinetic energy injection of
$\sim 10^{42}\:{\rm erg
\, s^{-1}}$, which is very similar to our estimate above.

It is interesting to compare this result to the energy input by
dynamical friction. However, it should be noted that our simulations
do not include the effect of dynamical friction (DF) since self-gravity is
not included.
\begin{equation}
  \left.
  \frac{{\rm d}E_{\rm gal}}{{\rm d} t}
  \right|_{\rm DF}
  =
  M_{\rm gal} v 
  \left.
  \frac{{\rm d}v}{{\rm d} t}
  \right|_{\rm DF}
  =
  -
  \frac{ 4 \pi \ln\Lambda \, G^2 \rho_{\rm ICM} M_{\rm gal}^2}{v}
  ,
\end{equation}
where $\ln\Lambda$ is the logarithm of the ratio of maximum and
minimum impact parameters and $M_{\rm gal}$ is the mass of the
galaxy. Using the same values as above, we obtain an energy loss of
$\sim 6\times10^{42}\:{\rm erg \, s^{-1}}$, which is somewhat above
the values derived for ram pressure heating. However, it is not clear
how efficiently the energy is converted into heat. The accelerated gas
may lead to propagating waves in the ICM that still need to be
dissipated. In contrast, in the wake behind the galaxy, the ICM
develops small-scale turbulence, which will dissipate locally.

Moreover, ram pressure heating scales with $A_{\rm
proj}$, whereas dynamical friction heating scales with
$M_{\rm gal}^2$. Hence, for the large number of small galaxies
ram pressure heating may prevail.

Our simulations indicate that a supersonic galaxy stirs the ICM much
more than a subsonically moving galaxy. Hence, gas that cools
significantly below the local galaxy velocity dispersion will be
heated more efficiently.
 
Compared to ICM temperatures, the ISM lost from galaxies by ram pressure
stripping is cold ($\sim 10^4$ K) and thus might be considered an energy sink
for the ICM. However, the energy that becomes available by decelerating this
stripped gas can be used for its heating. Given that cluster galaxies move
with typical velocities comparable to the ICM sound speed, this energy
approximately balances the energy needed to heat the stripped ISM to ICM
temperatures.


\section{Summary}
%
We ran high-resolution 3D hydrodynamical simulations of ram pressure stripping
of a massive disc galaxy focussing on the evolution of the galaxy's wake. We
combined the following parameters: a medium and a high ram pressure, one
subsonic and one supersonic ICM velocity, inclinations of $30\degree$ and
$75\degree$. We traced the tail to a distance of $260\Kpc$ behind the galaxy.

In our simulations, the wakes reached typical widths of $\sim 60$ to $100\Kpc$
at distances of $100\Kpc$ behind the galaxy. The widening proceeds
systematically with opening angles $\lesssim 23\degree$ in subsonic
runs. Depending on the projection angle, subsonic wakes can show oscillations
similar to von Karman vortex streets. The wakes in the supersonic cases also
show widening, but in an irregular fashion. 

We compared our simulations with the observed tail of NGC~4388. We could
explain some features, but our simulated tails are wider than this observed
one. We discussed possible reasons for this difference. 

Finally, we provided suggestions how our simulation results could be used in
cluster simulations that aim at predicting the evolution of the metal
distribution in the ICM due to ram pressure stripping. 


\section*{Acknowledgements}
We acknowledge the support by the DFG grant BR 2026/3 within the Priority
Programme ``Witnesses of Cosmic History'' and the supercomputing grants NIC
1927 and 1658 at the John-Neumann Institut at the Forschungszentrum J\"ulich.
The simulations were produced with STELLA, the LOFAR BlueGene/L
System. Special thanks for support go to H.~Falcke.

The results presented were produced using the FLASH code, a product of the DOE
ASC/Alliances-funded Center for Astrophysical Thermonuclear Flashes at the
University of Chicago.


%
\bibliographystyle{mn2e}
\bibliography{%
../../BIBLIOGRAPHY/theory_simulations,%
../../BIBLIOGRAPHY/hydro_processes,%
../../BIBLIOGRAPHY/numerics,%
../../BIBLIOGRAPHY/observations_general,%
../../BIBLIOGRAPHY/observations_clusters,%
../../BIBLIOGRAPHY/observations_galaxies,%
../../BIBLIOGRAPHY/galaxy_model,%
../../BIBLIOGRAPHY/gas_halo,%
../../BIBLIOGRAPHY/icm_conditions,%
../../BIBLIOGRAPHY/clusters,%
../../BIBLIOGRAPHY/else}

\begin{thebibliography}{}

\bibitem[\protect\citeauthoryear{Abadi, Moore \& Bower}{Abadi
  et~al.}{1999}]{abadi99}
Abadi M.~G.,  Moore B.,    Bower R.~G.,  1999, \mnras, 308, 947

\bibitem[\protect\citeauthoryear{Acreman, Stevens, Ponman \& Sakelliou}{Acreman
  et~al.}{2003}]{acreman03}
Acreman D.~M.,  Stevens I.~R.,  Ponman T.~J.,    Sakelliou I.,  2003, \mnras,
  341, 1333

\bibitem[\protect\citeauthoryear{Batchelor}{Batchelor}{2000}]{batchelor}
Batchelor G.~K.,  2000, An introduction to fluid dynamics.
Cambridge University Press

\bibitem[\protect\citeauthoryear{Binney \& Tremaine}{Binney \&
  Tremaine}{1987}]{binneytremaine}
Binney J.,  Tremaine S.,  1987, Galactic Dynamics.
Princeton University Press, Princeton, New Jersey

\bibitem[\protect\citeauthoryear{Burkert}{Burkert}{1995}]{burkert95}
Burkert A.,  1995, \apjl, 447, 25

\bibitem[\protect\citeauthoryear{Cayatte, Kontanyi, Balkowski \& van
  Gorkom}{Cayatte et~al.}{1994}]{cayatte94}
Cayatte V.,  Kontanyi C.,  Balkowski C.,    van Gorkom J.~H.,  1994, \aj, 107,
  1003

\bibitem[\protect\citeauthoryear{Cayatte, van Gorkom \& Kotanyi}{Cayatte
  et~al.}{1990}]{cayatte90}
Cayatte V.,  van Gorkom J.~H.,    Kotanyi C.,  1990, \aj, 100, 604

\bibitem[\protect\citeauthoryear{Davies, Minchin, Sabatini, {van Driel}, Baes,
  Boyce, {de Blok}, Disney, Evans, Kilborn, Lang, Linder, Roberts \&
  Smith}{Davies et~al.}{2004}]{davies04a}
Davies J.,  Minchin R.,  Sabatini S.,  {van Driel} W.,  Baes M.,  Boyce P.,
  {de Blok} W.~J.~G.,  Disney M.,  Evans R.,  Kilborn V.,  Lang R.,  Linder S.,
   Roberts S.,    Smith R.,  2004, \mnras, 349, 922

\bibitem[\protect\citeauthoryear{Domainko, Mair, Kapferer, {van Kampen},
  Kronberger, Schindler, Kimeswenger, Ruffert \& Mangete}{Domainko
  et~al.}{2005}]{domainko05}
Domainko W.,  Mair M.,  Kapferer W.,  {van Kampen} E.,  Kronberger T.,
  Schindler S.,  Kimeswenger S.,  Ruffert M.,    Mangete O.,  2005, \aap, ??,
  ??

\bibitem[\protect\citeauthoryear{{Faltenbacher}, {Kravtsov}, {Nagai} \&
  {Gottl{\"o}ber}}{{Faltenbacher} et~al.}{2005}]{faltenbacher05}
{Faltenbacher} A.,  {Kravtsov} A.~V.,  {Nagai} D.,    {Gottl{\"o}ber} S.,
  2005, \mnras, 358, 139

\bibitem[\protect\citeauthoryear{{Fryxell}, {Olson}, {Ricker}, {Timmes},
  {Zingale}, {Lamb}, {MacNeice}, {Rosner}, {Truran} \& {Tufo}}{{Fryxell}
  et~al.}{2000}]{fryxell00}
{Fryxell} B.,  {Olson} K.,  {Ricker} R.,  {Timmes} F.~X.,  {Zingale} M.,
  {Lamb} D.~Q.,  {MacNeice} P.,  {Rosner} R.,  {Truran} J.~W.,    {Tufo} H.,
  2000, \apjs, 131, 273

\bibitem[\protect\citeauthoryear{Gavazzi, Boselli, Mayer, Iglesias-Paramo,
  V\'ilchez \& Carrasco}{Gavazzi et~al.}{2001}]{gavazzi01}
Gavazzi G.,  Boselli A.,  Mayer L.,  Iglesias-Paramo J.,  V\'ilchez J.~M.,
  Carrasco L.,  2001, \apj, 563, 23L

\bibitem[\protect\citeauthoryear{Gavazzi, Contursi, Carrasco, Boselli,
  Kennicutt, Scodeggio \& Jaffe}{Gavazzi et~al.}{1995}]{gavazzi95}
Gavazzi G.,  Contursi A.,  Carrasco L.,  Boselli A.,  Kennicutt R.,  Scodeggio
  M.,    Jaffe W.,  1995, \aap, 304, 325

\bibitem[\protect\citeauthoryear{Hernquist}{Hernquist}{1993}]{hernquist93}
Hernquist L.,  1993, \apjs, 86, 389

\bibitem[\protect\citeauthoryear{Kenney \& Koopmann}{Kenney \&
  Koopmann}{1999}]{kenney99}
Kenney J.~D.~P.,  Koopmann R.~A.,  1999, \aj, 117, 181

\bibitem[\protect\citeauthoryear{Kenney \& Koopmann}{Kenney \&
  Koopmann}{2001}]{kenney01}
Kenney J.~D.~P.,  Koopmann R.~A.,  2001, in Hibbard J.~E.,  Rupen M.~P.,   van
  Gorkom J.~J.,  eds, Gas and Galaxy Evolution Vol.~240 of ASP Conf. Ser.,
  Environmental effects on gas and star formation in virgo cluster spiral and
  peculiar galaxies.
ASP, San Francisco, p.~577

\bibitem[\protect\citeauthoryear{Kenney, {van Gorkom} \& Vollmer}{Kenney
  et~al.}{2004}]{kenney04}
Kenney J.~D.~P.,  {van Gorkom} J.~H.,    Vollmer B.,  2004, \aj, 127, 3361

\bibitem[\protect\citeauthoryear{Kim, {El-Zant} \& {Kamionkowski}}{Kim
  et~al.}{2005}]{kim05}
Kim W.,  {El-Zant} A.~A.,    {Kamionkowski} M.,  2005, \apj, 632, 157

\bibitem[\protect\citeauthoryear{Koopmann, Haynes \& Catinella}{Koopmann
  et~al.}{2005}]{koopmann05}
Koopmann R.~A.,  Haynes M.~P.,    Catinella B.,  2005, \aj{}, in press,
  astro-ph/0510374

\bibitem[\protect\citeauthoryear{Koopmann \& Kenney}{Koopmann \&
  Kenney}{1998}]{koopmann98}
Koopmann R.~A.,  Kenney J.~D.~P.,  1998, \apj, 497, 75L

\bibitem[\protect\citeauthoryear{Koopmann \& Kenney}{Koopmann \&
  Kenney}{2004a}]{koopmann04b}
Koopmann R.~A.,  Kenney J.~D.~P.,  2004a, \apj, 613, 866

\bibitem[\protect\citeauthoryear{Koopmann \& Kenney}{Koopmann \&
  Kenney}{2004b}]{koopmann04a}
Koopmann R.~A.,  Kenney J.~D.~P.,  2004b, \apj, 613, 851

\bibitem[\protect\citeauthoryear{Marcolini, Brighenti \& A.D'Ercole}{Marcolini
  et~al.}{2003}]{marcolini03}
Marcolini A.,  Brighenti F.,    A.D'Ercole 2003, \mnras, 345, 1329

\bibitem[\protect\citeauthoryear{Miyamoto \& Nagai}{Miyamoto \&
  Nagai}{1975}]{miyamoto75}
Miyamoto M.,  Nagai R.,  1975, \pasj, 27, 533

\bibitem[\protect\citeauthoryear{Mori \& Burkert}{Mori \&
  Burkert}{2000}]{mori00}
Mori M.,  Burkert A.,  2000, \apj, 538, 559

\bibitem[\protect\citeauthoryear{Oosterloo \& {van Gorkom}}{Oosterloo \& {van
  Gorkom}}{2005}]{oosterloo05}
Oosterloo T.,  {van Gorkom} J.,  2005, \aap, 437, L19

\bibitem[\protect\citeauthoryear{Ostriker}{Ostriker}{199}]{ostriker99}
Ostriker E.~C.,  199, \apj, 513, 252

\bibitem[\protect\citeauthoryear{Quilis, Moore \& Bower}{Quilis
  et~al.}{2000}]{quilis00}
Quilis V.,  Moore B.,    Bower R.,  2000, Science, 288, 1617

\bibitem[\protect\citeauthoryear{Reynolds, {McKernan}, Fabian, Stone \&
  Vernaleo}{Reynolds et~al.}{2005}]{reynolds05}
Reynolds C.~S.,  {McKernan} B.,  Fabian A.~C.,  Stone J.~M.,    Vernaleo J.~C.,
   2005, \mnras, 357, 242

\bibitem[\protect\citeauthoryear{Roediger \& Br\"uggen}{Roediger \&
  Br\"uggen}{2006}]{roediger06}
Roediger E.,  Br\"uggen M.,  2006, \mnras, submitted, astro-ph/0512365

\bibitem[\protect\citeauthoryear{Roediger \& Hensler}{Roediger \&
  Hensler}{2005}]{roediger05}
Roediger E.,  Hensler G.,  2005, \aap, 433, 875

\bibitem[\protect\citeauthoryear{Schindler, Kapferer, Domainko, Mair, {van
  Kampen}, Kronberger, Kimeswenger, Ruffert, Mangete \&
  Breitschwerdt}{Schindler et~al.}{2005}]{schindler05}
Schindler S.,  Kapferer W.,  Domainko W.,  Mair M.,  {van Kampen} E.,
  Kronberger T.,  Kimeswenger S.,  Ruffert M.,  Mangete O.,    Breitschwerdt
  D.,  2005, \aap, 435, L25

\bibitem[\protect\citeauthoryear{Schulz \& Struck}{Schulz \&
  Struck}{2001}]{schulz01}
Schulz S.,  Struck C.,  2001, \mnras, 328, 185

\bibitem[\protect\citeauthoryear{Solanes, Manrique, Garc\'ia-G\'omez,
  Gonz\'ales-Casando, Giovanelli \& Haynes}{Solanes et~al.}{2001}]{solanes01}
Solanes J.~M.,  Manrique A.,  Garc\'ia-G\'omez C.,  Gonz\'ales-Casando G.,
  Giovanelli R.,    Haynes M.~P.,  2001, \apj, 548, 97

\bibitem[\protect\citeauthoryear{Stevens, Acreman \& Ponman}{Stevens
  et~al.}{1999}]{stevens99}
Stevens I.~R.,  Acreman D.~M.,    Ponman T.~J.,  1999, \mnras, 310, 663

\bibitem[\protect\citeauthoryear{Takizawa}{Takizawa}{2005}]{takizawa05a}
Takizawa M.,  2005, \apj, 629, 791

\bibitem[\protect\citeauthoryear{Vollmer, Beck, Kenney \& {van Gorkom}}{Vollmer
  et~al.}{2004}]{vollmer04a}
Vollmer B.,  Beck R.,  Kenney J.~D.~P.,    {van Gorkom} J.~H.,  2004, \aj, 127,
  3375

\bibitem[\protect\citeauthoryear{Vollmer, Braine, Balkowski, Cayatte \&
  Duschl}{Vollmer et~al.}{2001}]{vollmer01}
Vollmer B.,  Braine J.,  Balkowski C.,  Cayatte V.,    Duschl W.~J.,  2001,
  \aap, 374, 824

\bibitem[\protect\citeauthoryear{Vollmer, Cayatte, Balkowski \& Duschl}{Vollmer
  et~al.}{2001}]{vollmer01a}
Vollmer B.,  Cayatte V.,  Balkowski C.,    Duschl W.~J.,  2001, \apj, 561, 708

\bibitem[\protect\citeauthoryear{Vollmer, Cayatte, Boselli, Balkowski \&
  Duschl}{Vollmer et~al.}{1999}]{vollmer99}
Vollmer B.,  Cayatte V.,  Boselli A.,  Balkowski C.,    Duschl W.~J.,  1999,
  \aap, 349, 411

\bibitem[\protect\citeauthoryear{Vollmer \& Huchtmeier}{Vollmer \&
  Huchtmeier}{2003}]{vollmer03a}
Vollmer B.,  Huchtmeier W.,  2003, \aap, 406, 427

\end{thebibliography}

\bsp

\label{lastpage}

\end{document}